\pdfoutput=1
\documentclass[aps, prd,twocolumn,showpacs, floatfix, nofootinbib,
preprintnumbers
]{revtex4}
\usepackage{amsmath}
\usepackage{amssymb}
\usepackage{epsfig}
\usepackage{graphicx}
\usepackage{color}
\usepackage{dsfont}
\usepackage{array} 

\newcommand{\be}{\begin{equation}}
\newcommand{\ee}{\end{equation}}
\newcommand{\bdm}{\begin{displaymath}}
\newcommand{\edm}{\end{displaymath}}
\newcommand{\bea}{\begin{eqnarray}}
\newcommand{\eea}{\end{eqnarray}}
\newcommand{\nn}{\nonumber}

\newcommand{\unit}{\mathds{1}}

\newcommand{\GeV}{\,\mathrm{GeV}}

\begin{document}
%----------------------------------------------------------------------------------
\title{Proton lifetime in the minimal $SO(10)$ GUT and its implications for the LHC}
\preprint{}
\pacs{12.10.-g, 12.10.Kt, 14.80.-j}
%----------------------------------------------------------------------------------
\author{Helena Kole\v{s}ov\'a}\email{helena.kolesova@fjfi.cvut.cz}
\affiliation{Institute of Particle and Nuclear Physics,
Faculty of Mathematics and Physics,
Charles University in Prague, V Hole\v{s}ovi\v{c}k\'ach 2,
180 00 Praha 8, Czech Republic \\ and \\Faculty of Nuclear Sciences and Physical Engineering, Czech Technical University in Prague, B\v{r}ehov\'a 7, 115 19 Praha 1, Czech Republic
}
\author{Michal Malinsk\'{y}}\email{malinsky@ipnp.troja.mff.cuni.cz}
\affiliation{Institute of Particle and Nuclear Physics,
Faculty of Mathematics and Physics,
Charles University in Prague, V Hole\v{s}ovi\v{c}k\'ach 2,
180 00 Praha 8, Czech Republic}
%----------------------------------------------------------------------------------
\begin{abstract}
We review the current status of the minimal non-supersymmetric $SO(10)$ grand unified theory and perform a detailed next-to-leading-order analysis of the gauge unification and proton lifetime constraints on the part of its parameter space supporting a ZeV-scale color sextet scalar. This, together with a TeV-scale color octet studied in detail in a preceding work, represents one of the two minimally fine-tuned settings compatible with all the relevant consistency and phenomenology limits. Both these scenarios can be  extensively tested at the future megaton-scale proton-decay facilities. On top of that, the light octet solution can be accessible in the TeV-scale collider searches.  
\end{abstract}
%----------------------------------------------------------------------------------
\maketitle
%%%%%%%%%%%%%%%%%%%%%%%%%%%%%%%%%%%%%%
\section{Introduction}
%%%%%%%%%%%%%%%%%%%%%%%%%%%%%%%%%%%%%%
With  the advent of the new multi-kiloton-scale neutrino experiments discussed intensively in the last years such as Hyper-Kamiokande (HK)~\cite{Abe:2011ts}, LBNE~\cite{Akiri:2011dv} or LENA~\cite{Oberauer:2005kw} there is a good chance that a new type of  beyond-Standard-Model (BSM) signals would be revealed in the near future, possibly, in the next decade. 
In this respect, the recent decision of the Japanese Science Council to list the HK proposal among the top 27~ventures of the ``Japanese Master Plan of Large Projects'' is clearly a great step towards these goals.  

However, the -- by many expected -- CP violation in the lepton sector associated with the so-called Dirac phase of the leptonic mixing matrix is not the only new physics  such a machine can shed light on. Together with the number of ongoing searches for the neutrinoless double beta decay there is a good chance to get soon an answer to a yet more fundamental question of whether the baryon and/or lepton numbers, accidental global symmetries of the Standard Model (SM) Lagrangian, are respected by the BSM physics.

The prominent signal of baryon number violation (BNV) accessible, at least in principle, by these machines  is the hypothetical instability of protons. The current best limits on proton lifetime from the Super-Kamiokande (SK) experiment~\cite{Nishino:2012ipa} reach, at 90\% C.L., $8.2\times 10^{33}$ years in the $p^{+}\to \pi^{0} e^{+}$ channel and up to $2.3\times 10^{33}$ for $p^{+}\to K^{+}\nu$, cf.~\cite{Kobayashi:2005pe}. Entering the megaton-scale range with the water-Cherenkov technology of the HK (and the 30 kiloton ballpark with the liquid argon TPC at LBNE) makes it possible to improve the current sensitivity by at least one order of magnitude in (not only) these principal channels.

In a sharp contrast to the steady (though slow) progress in experiment, on the theory side the proton instability has never been addressed in full consistency at a better than the leading-order (LO) accuracy level. Paradoxically, even after several decades of continuous efforts the existing proton lifetime estimates are still systematically plagued by several-orders-of-magnitude uncertainties. Although the main source of these errors -- the large uncertainty in the LO GUT-scale determination -- can be constricted by focusing on branching ratios rather than at the absolute proton decay width, the best such approach can provide is just a limited discrimination between wide classes of models with similar flavor structure.

There seem to be several historical reasons why this happened to be so. The initial studies of the gauge coupling unification~\cite{Georgi:1974yf,Goldman:1980ah,Langacker:1980js} suffered from the lack of reliable input data which, in turn, hindered the determination of the GUT scale with better than few-orders-of-magnitude precision.
The new data arriving in 1980's from the CERN's SPS and Fermilab's Tevatron (and later from the LEP) refuted the original minimal SU(5) model~\cite{Georgi:1974sy} as unable to account for the measured value of the weak mixing angle. This, in turn, was re-interpreted as a hint for a TeV-scale supersymmetry (SUSY) so the subsequent failure of the minimal non-SUSY $SO(10)$ GUT~\cite{Yasue:1980fy,Anastaze:1983zk,Babu:1984mz} due to its notorious trouble with tachyonic instabilities was not acknowledged too much; indeed, since about mid 1980's there was already the new SUSY GUT paradigm.

Unfortunately, as attractive as the low-energy SUSY idea was in its early days, it was conceptually ruinous for any conclusive testability of unified models because, without any solid information on the SUSY spectra, the proton decay rate (dominated in SUSY by $d=5$ loop diagrams including squarks, sleptons, gauginos and higgsinos) was incalculable. The salvation was expected from the LHC which was generally assumed to find SUSY states in the TeV domain. Thus, for the last twenty years the field was in a slightly ``schizophrenic'' situation in which the unification idea was (in a rudimentary form) implemented in most of the ``big'' theoretical endeavors like, for instance, detailed SUSY phenomenology studies but, at the same time, its testability was obscured by the very concept it was supposed to reinforce.

These concerns became yet more acute in the last decade when a strong tension among the rigid flavor structures of the minimal renormalizable SUSY $SU(5)$ and $SO(10)$ GUTs~\cite{Murayama:2001ur,Bajc:2002bv,Bajc:2002pg,Aulakh:2005mw,Bertolini:2006pe} and the new proton lifetime and neutrino data was revealed. On top of that, there are no signs of salvation emerging in the latest LHC SUSY searches and, as a matter of fact, there is no reason to expect any spectacular proliferation of SUSY states in the high-energy phase of the LHC either.

This, all together, leads to the rather unsatisfactory current picture where the typical errors quoted (sometimes even not) in the existing proton lifetime studies  stretch easily up to four or five orders of magnitude.  As good as it may have been several decades ago this becomes a real issue when the cost of the next-generation machines (capable of boosting the sensitivity by ``only'' about a factor of ten) becomes comparable to that of the most expensive  particle physics assets such as the LHC.
From this perspective, the megaton-scale proton decay facilities such as the HK would benefit enormously from a firm and robust next-to-leading-order (NLO) proton lifetime estimates with theory errors contained within their order-of-magnitude-wide improvement windows. 

However, such fully consistent NLO studies call for an unprecedented level of complexity and, in most cases,  they turn out to be even impossible.
First, at the NLO level, the unification scale should  be determined via a thorough two-loop renormalization group analysis. This is certainly no problem as far as the $\beta$-functions are concerned; these are nowadays routinely calculated up to three-loops (and even more). However, there is no point in dealing with the resulting evolution equations without a complete account of the relevant threshold corrections at the corresponding accuracy level.  Hence, a very important ingredient of any potentially consistent analysis is a good information about not only the light but, namely, the heavy (GUT-scale) part of the theory spectrum. This issue, however, is often ignored in bottom-up studies in which simplified assumptions about the shape of the heavy spectrum are invoked, thus inflicting an irreducible theoretical uncertainty  in the unification scale $M_{G}$ comparable in size to the two-loop $\beta$-function effects.  

Moreover, there are classes of uncertainties that, in most cases, ruin the reliability of even the relatively simple two-loop RG running studies. These have to do namely with the proximity of $M_{G}$ to the Planck scale $M_{P}$ and, thus, the alleged sensitivity of the GUT-scale physics to the quantum gravity effects (parametrized, at the effective level, by the $M_{G}/M_{P}$-suppressed operators). These so called ``gravity smearing effects'' \cite{Larsen:1995ax,Veneziano:2001ah,Calmet:2008df,Dvali:2007hz,Chakrabortty:2008zk} usually amount to about a few $\%$-level uncertainty in the GUT-scale matching conditions which, however, translate to a significant error in $M_{G}$ due to the relative ``shallowness'' of the intersection pattern of the logarithmically evolving gauge couplings, again at the level of a typical 2-loop $\beta$-function effect.  
 
Second, the generally complicated structure of the BNV currents coupled to the colored extra scalars and vectors calls for a detailed understanding of the  flavor pattern of the GUT models under consideration. This, at the NLO level, amounts to getting a good grip onto their spectra as well as onto the related RG evolution and its matching to the existing low-energy inputs which applies not only to the gauge and Yukawa couplings but also to the relevant  d=6 effective BNV operators, cf.~\cite{Nath:2006ut,Buras:1977yy,Ellis:1979hy,Wilczek:1979hc}.

Although most of these issues have been extensively discussed in the literature, a patient and systematic synthesis of all these aspects is still missing. Needless to say, this can be efficiently attempted only in a very limited class of the simplest models where the main theoretical bias, namely, the ``gravity smearing'', is under control.  This, however, is almost never the case; alas, this turns out to be quite hopeless in SUSY GUTs, see, e.g.,~\cite{Pati:2005yp,Murayama:2001ur,Dutta:2004zh}. In view of this, it is not surprising that accurate proton lifetime estimates never became part of the mainstream and it was even less so in the pre-LHC era when the SUSY paradigm was prevalent.

In this study we would like to review the status and prospects of one of the rare exceptions to this ``NLO no-go'', namely, the minimal non-supersymmetric $SO(10)$ GUT~\cite{Yasue:1980fy,Anastaze:1983zk,Babu:1984mz}. Remarkably enough, in this scenario the trouble with the leading Planck-scale effects in the GUT-scale determination is alleviated by the fact that the 45-dimensional $SO(10)$ adjoint representation $\Phi$ responsible for the GUT symmetry breaking can not couple to the pair of the gauge field strength tensors at the $d=5$ level (i.e.,  $G^{a}_{\mu\nu}\Phi^{ab}G^{b\mu\nu}=0$) and, thus, the ``gravity smearing'' effects  are absent at the leading order~\cite{Nath:2006ut}. Unfortunately, this model was left aside for many years due to the aforementioned problem with the tachyonic instabilities (perhaps even more probably due to its non-SUSY nature) and it was  revived only recently \cite{Bertolini:2009es} as a GUT that is entirely consistent at the {\em quantum} level. 

Besides recapitulating the salient features of the model and the general constraints implied by the requirements of gauge unification and vacuum stability, we focus namely on the interplay between its different low-energy aspects, in particular, the non-observation of proton decay at the SK and the absence of light exotics at the LHC.   
In doing so, we complement the existing two-loop analysis of the setting featuring an accidentally light scalar color octet~\cite{Bertolini:2013vta}  with a detailed two-loop study of the second potentially realistic minimally-finetuned option identified in~\cite{Bertolini:2012im} with an accidentally light color sextet at about $10^{12}$ GeV. As we shall argue, this scenario, as fine as it looks at the one-loop level, would become strongly constrained at two loops if there would be no proton decay seen at the HK. Hence, in the vast majority of the available parameter space (defining  ``available'' as ``consistent with observation'' and barring for the moment the arbitrariness in choosing a measure on it) the model practically admits only one type of solutions conforming all the basic phenomenological requirements, namely, those featuring the very light color octet, which, in turn, implies a signal observable either at the HK or at the multi-TeV hadronic colliders such as the LHC or its near-future successors. 

The work is organized as follows:  In section II we begin with a brief recapitulation of the salient features of the minimal SO(10) GUT including a simplified account of the quantum effects necessary for the technical stabilization of its non-SU(5)-like vacua. Section III is devoted to the discussion of what we consider to be the minimal consistent approach to any numerical analysis in this framework. The main guiding principle here is the overall consistency and generality of the obtained results; the latter can be trivially translated to the requirement of the minimum number of fine-tunings (assuming ``flat'' parameter-space measure). In this approach, all the potentially realistic areas of the parameter space are covered up to the subsets of zero measure. To that end, we recapitulate the existing results for the two classes of known minimally-fine-tuned solutions, namely, those featuring a near-TeV-scale colored octet transforming like $(8,2,+\tfrac{1}{2})$ under the SM and an intermediate-scale colored sextet with the SM quantum numbers $(6,3,+\tfrac{1}{3})$. Motivated by the significant quantitative change in the   behavior of the octet solution observed at the transition from the LO to the NLO level, cf.~\cite{Bertolini:2013vta}, in Section IV we extend the existing LO sextet analysis to the same NLO level. In doing so, we reveal a strong correlation between the position of the GUT scale favored by the relevant NLO solutions and the typical value of the seesaw scale which, from the perspective of the detailed flavor sector fits, turns out to be rather low. Finally, in Section V we comment in brief on the overall viability of the minimal setting and recapitulate its distinctive phenomenological features.  

%%%%%%%%%%%%%%%%%%%%%%%%%%%%%%%%%%%%%%
\section{The minimal $SO(10)$ GUT}
%%%%%%%%%%%%%%%%%%%%%%%%%%%%%%%%%%%%%%
Since the gauge interactions of the matter fields (accommodated, as usual, in three copies of the 16-dimensional $SO(10)$ spinors $\psi_{i}$) are fully specified by the minimal coupling principle, the renormalizable grand unified models are essentially defined by the structure of their scalar sector and the corresponding Yukawa Lagrangian. 

As for the former,  the minimal $SO(10)$ model of our interest contains the anticipated 45-dimensional adjoint scalar representation $\phi$ responsible for the spontaneous breaking of the GUT symmetry down to one of its rank-5 subgroups (see Sect.~\ref{sect:potvac}) while the rank reduction is triggered by the complex 126-dimensional 5-index antisymmetric tensor $\Sigma$. This choice is motivated namely by the need to generate neutrino masses in the sub-eV ballpark: since the gauge unification constraints typically place the rank-reducing dynamics (i.e., the $B-L$ breaking scale) to about $10^{10-12}$ GeV~\cite{Bertolini:2009qj} the corresponding VEV $\sigma$ should enter the RH neutrino masses linearly, otherwise the effective RH neutrino mass scale would be further suppressed (typically by $\sigma/M_{Pl}$ to some positive integer power) and, hence, the seesaw-induced light neutrinos will tend to be far too heavy to conform the cosmological constraints. This, in the current scenario, is achieved by the renormalizable Yukawa coupling of the type ${\cal L}\ni Y^{\Sigma}_{ij}\psi_{i}\psi_{j}\Sigma^{*}$. Needless to say, besides $\Sigma$ there should be at least one more ``Yukawa-active'' scalar at play (such as, for instance, a 10-dimensional SO(10) vector)  in order to accommodate the quark and lepton masses and mixing data; however,  since what follows is largely independent on these details we shall not elaborate on the specific structure of the Yukawa sector here.

\subsection{The effective SO(10)-breaking scalar potential\label{sect:potvac}}
The renormalizable tree-level scalar potential of the minimal $SO(10)$ model under consideration reads
\be
\label{scalpotgen}
V = V_{45} + V_{126} + V_{\rm mix} \, ,
\ee
where
\bea
\label{V45}
V_{45} &=& - \frac{\mu^2}{2} (\phi \phi)_0 +\!\frac{a_0}{4} (\phi \phi)_0 (\phi \phi)_0 +\! \frac{a_2}{4} (\phi \phi)_2 (\phi \phi)_2, \\ \nn \\
\label{V126}
V_{126} &=&  - \frac{\nu^2}{5!} (\Sigma \Sigma^*)_0\\
& +& \frac{\lambda_0}{(5!)^2} (\Sigma \Sigma^*)_0 (\Sigma \Sigma^*)_0
+ \frac{\lambda_2}{(4!)^2} (\Sigma \Sigma^*)_2 (\Sigma \Sigma^*)_2\nn\\
& + & \frac{\lambda_4}{(3!)^2(2!)^2} (\Sigma \Sigma^*)_4 (\Sigma \Sigma^*)_4
+ \frac{\lambda'_{4}}{(3!)^2} (\Sigma \Sigma^*)_{4'} (\Sigma \Sigma^*)_{4'} \nn \\ \nn \\
&+& \frac{\eta_2}{(4!)^2} (\Sigma \Sigma)_2 (\Sigma \Sigma)_2
+ \frac{\eta_2^*}{(4!)^2} (\Sigma^* \Sigma^*)_2 (\Sigma^* \Sigma^*)_2 \, , \nn \\ \nn \\
\label{V45126}
V_{\rm mix} &=& \frac{i \tau}{4!} (\phi)_2 (\Sigma \Sigma^*)_2
+ \frac{\alpha}{2 \cdot 5!} (\phi \phi)_0 (\Sigma \Sigma^*)_0 \\
%+ \beta_2 (\phi \phi)_2 (\Sigma \Sigma^*)_2
&+& \frac{\beta_4}{4 \cdot 3!} (\phi \phi)_4 (\Sigma \Sigma^*)_4
+ \frac{\beta'_{4}}{3!} (\phi \phi)_{4'} (\Sigma \Sigma^*)_{4'} \nn \\ \nn \\
&+& \frac{\gamma_2}{4!} (\phi \phi)_2 (\Sigma \Sigma)_2
+ \frac{\gamma_2^*}{4!} (\phi \phi)_2 (\Sigma^* \Sigma^*)_2 \,\nn .
\eea
Here the subscripts of the round brackets denote different contractions of the fields within; let us also note that all couplings but $\eta_2$~and~$\gamma_2$ are real. For more details the reader is referred to~\cite{Bertolini:2012im}.

The high-scale symmetry breaking is triggered by the SM-singlet VEVs in $\phi$ and $\Sigma$ that we shall denote by
\bea
& \langle(1,1,1,0)_{\phi}\rangle\equiv \omega_{BL},\,\, \langle(1,1,3,0)_{\phi}\rangle\equiv \omega_R, & \label{vevs}\\ 
& \langle(1,1,3,+2)_{\Sigma}\rangle\equiv \sigma, &\nn
\eea
where the relevant components are classified with respect to the $SU(3)_c\otimes SU(2)_L\otimes SU(2)_R\otimes U(1)_{BL}$ subgroup of the $SO(10)$.
Without loss of generality we may assume $\omega_{BL}$ and $\omega_R$ to be real; $\sigma$ can be also made real by a phase redefinition of $\Sigma$. 

As usual, the residual symmetry depends on the specific configuration of the VEVs~(\ref{vevs}).
Taking, for the moment, $\sigma=0$, several interesting limits can be distinguished:
\begin{align}
&\omega_R=0,\omega_{BL}\neq 0: &3_c2_L2_R 1_{BL},\nonumber\\
&\omega_R\neq 0,\omega_{BL}= 0: &4_C 2_L1_R,\nonumber\\
&\omega_R\neq 0,\omega_{BL}\neq 0: &3_c2_L1_R 1_{BL},\label{regimes}\\
&\omega_R=-\omega_{BL}\neq 0: &\mathrm{flipped }\,5'1_{Z'},\nonumber\\
&\omega_R=\omega_{BL}\neq 0: &\mathrm{standard }\,5 1_{Z},\nonumber
\end{align}
where the acronyms on the right-hand side (RHS) of each line denote (in a self-explanatory notation) the relevant little group. To this end, $5 1_{Z}$ and $5'1_{Z'}$ represent the two different embeddings of the SM hypercharge into the $SU(5)\times U(1)$ subgroup of $SO(10)$ usually referred to as the ``standard'' and the ``flipped'' $SU(5)$ scenarios, respectively (for further details see, e.g., \cite{Bertolini:2009es}). A nonzero $\sigma$ breaks all these intermediate symmetries down to the SM group with the only exception being the last case where the $SU(5)$ symmetry remains unbroken.

The tree-level scalar masses may be computed from the potential \eqref{scalpotgen} (for a complete list of the relevant formulae see Appendix B of \cite{Bertolini:2013vta}). Remarkably enough, such a scalar spectrum suffers from a notorious tachyonic instability~\cite{Yasue:1980fy,Anastaze:1983zk,Babu:1984mz} having to do, namely, with the masses of the fields with the SM quantum numbers $(1,3,0)$ and $(8,1,0)$:
\bea
\label{treemass130}
M^2 (1,3,0) &=& 2 a_2 (\omega_{BL} - \omega_R) (\omega_{BL} + 2 \omega_R) \, , \\
\label{treemass810}
M^2 (8,1,0) &=& 2 a_2 (\omega_R - \omega_{BL}) (\omega_R + 2 \omega_{BL}) \,.
\eea
It is straightforward to see that one of these always becomes negative unless $\omega_{R}$ and $\omega_{BL}$ are aligned along the ``approximate $5'1_{Z'}$ direction''
\be\label{danger}
-2\leq \frac{\omega_{BL}}{\omega_{R}}\leq -\frac{1}{2}\;\;\text{ with }\;\;a_{2}<0.
\ee
This, however, is incompatible with the gauge running constraints because, due to the proximity of the unification point, the corresponding $5' 1_{Z'}$-like symmetry breaking pattern resembles that of the of the long-ago refuted  minimal $SU(5)$ theory.

\subsection{The one-loop vacuum\label{oneloop}}
It is well known that these tachyonicity/vacuum instability issues may be resolved at the quantum level~\cite{Bertolini:2009es}. The point is that the extremely simplistic form of the two critical relations (\ref{treemass130}) and (\ref{treemass810}) can be traced back to the particular algebraic structure of the scalar potential (\ref{scalpotgen}) that prevents some of its couplings from entering these mass formulae at the tree level due to the pseudo-Goldstone nature of the corresponding fields. This degeneracy, however, is smeared at the loop level and, thus, there is much more room for arranging a tachyon-free scalar spectrum in the physically interesting regimes with $|\omega_{R}|\gg |\omega_{BL}|$ or  $|\omega_{BL}|\gg |\omega_{R}|$, i.e., those far from the dangerous $SU(5)$-like settings~(\ref{danger}). 

As an example, let us consider the gauge contributions to $M^2 (1,3,0)$ and $M^2 (8,1,0)$ that, at the leading loop level, are identical to those calculated in the simplified setting with the scalar $16$ in place of $126$, cf.~\cite{Bertolini:2009es}. The relevant one-loop formulae read:
\bea
\label{310onthevac}
M^2(1,3,0)&=&2 a_2 (\omega_{BL} - \omega_R) (\omega_{BL} + 2 \omega_R)\\
& & + \frac{g^4}{4\pi^2}
 \left(16 \omega _R^2+\omega _{BL} \omega _R+19 \omega _{BL}^2\right) +\ldots\,,
\nn\\[1ex]
\label{810onthevac}
M^2(8,1,0)&=&  2 a_2 (\omega_R - \omega_{BL}) (\omega_R + 2 \omega_{BL}) \\
& & +\frac{g^4 }{4\pi^2} 
\left(13 \omega _R^2+\omega _{BL} \omega _R+22 \omega _{BL}^2\right)+\ldots \nn,
\eea
where the ellipsis stand for $\beta^{2}$- and $\tau^{2}$-proportional terms polynomial in $\omega_R$ and $\omega_{BL}$ as well as for all the logarithmic terms. 
 It is clear that in order for the positive gauge corrections to overwhelm the potentially negative $a_{2}$-proportional terms it is sufficient to take $|a_{2}|$ small enough, typically in the few percent ballpark.  

Finally, let us note that, unlike for the simplified setting with $45\oplus16$ in the Higgs sector, only some of the undisplayed radiative corrections in (\ref{310onthevac}) and (\ref{810onthevac}) have been calculated so far (in particular, the SO(10) invariant term proportional to $\tau^{2}$, cf.~\cite{Bertolini:2013vta}); however, this issue should not affect the analysis below in any significant manner.
As it was argued in~\cite{Bertolini:2009es}, only the masses of the would-be-tachyonic fields  $(1,3,0)$ and $(8,1,0)$ may experience significant shifts due to quantum corrections; however, their possible effects in the relevant matching formulae (cf. Appendix~\ref{sect:matching}), are typically suppressed with respect to those of the other fields, e.g., the gauge degrees of freedom.
Hence, in the calculations below we stick to just the minimal set of loop corrections that suffice to tame the tachyonic instabilities (in particular, those calculated in~\cite{Bertolini:2013vta}). We verified explicitly that all our results are robust with respect to these theoretical uncertainties.

%%%%%%%%%%%%%%%%%%%%%%%%%%%%%%%%%%%%%%
\section{Consistent settings\label{sect:consistency}}
%%%%%%%%%%%%%%%%%%%%%%%%%%%%%%%%%%%%%%
\subsection{General considerations}
Let us begin with a list of basic constraints shaping the allowed patches of the parameter space supporting the consistent and potentially realistic settings in the  model of our interest. These have been previously discussed in great detail in~\cite{Bertolini:2012im} so here we shall just briefly recapitulate them.

\subsubsection{Theoretical consistency}

\paragraph{Perturbativity constraints.} Perturbativity is the primary principle we shall adhere to otherwise there is not much one can say quantitatively about the NLO structure of the theory. In particular, we shall assume that all couplings in the scalar potential (\ref{scalpotgen}) and also the Yukawa couplings are within the ${\cal O}(1)$  domain.

\paragraph{Non-tachyonicity of the scalar spectrum, local vacuum stability.} A negative eigenvalue of the scalar mass(squared)-matrix signals that the chosen field configuration is not a true vacuum of the model with all the unpleasant implications for the consistency of the broken-phase perturbation theory developed around such a setting.
Hence, the basic consistency requirement one should impose is that all scalar mass-squares calculated for a given field configuration should be positive for all non-Goldstone directions; this, in turn, ensures the local stability of the vacuum of the theory and a meaningful interpretation of its asymptotic states.  
Due to the rather complicated structure of the scalar potential (\ref{scalpotgen})  the discussion of the global stability of the electroweak vacuum will be left to a dedicated future study.

\paragraph{Gauge unification constraints.}
As we already mentioned,  in the non-SUSY context one generally needs to ``populate the  desert'' between the electroweak scale $M_{Z}$ and the GUT scale $M_{G}$ to some degree in order to conform the gauge unification constraints. As a matter of fact, this picture is also favored by the seesaw approach to the neutrino masses that, in its simplest incarnations, calls for a new scale in roughly the $10^{12-14}$ GeV ballpark. Note that, unlike in the bottom-up effective scenarios, this may be more difficult to achieve in the unified top-down approach due to the tight correlations in the Yukawa sector of GUTs that usually do not leave much room for tweaking; indeed, this is one of the generic issues plaguing the minimal supersymmetric SO(10) GUTs~\cite{Aulakh:2005mw,Bertolini:2006pe}.
From this perspective, the past RG results~\cite{Chang:1984qr,Deshpande:1992au,Deshpande:1992em,Bertolini:2009qj} based on the extended survival hypothesis (ESH)~\cite{delAguila:1980at,Mohapatra:1982aq,Aulakh:1999pz}  may be discouraging because they uniformly favor the seesaw-driving VEV $\sigma$ to be at around $10^{10}$ GeV which is rather far from the numbers above. On the other hand, these studies by their nature ignore all the details of the scalar sector and, thus, should be interpreted with care. In what follows we shall check that each of the accepted points in the parameter space yields a  consistent gauge unification pattern at two loops including the all-important matching effects \`a~la Weinberg~\cite{Weinberg:1980wa} and Hall~\cite{Hall:1980kf}.

\paragraph{Minimal number of fine-tunings\label{minimalfinetuning}.} 
Since, technically, there is no difference among pulling down the seesaw scale to be well below $M_{G}$ (i.e., performing a fine-tuning in the one-point function of the appropriate SM singlet field) or bringing down a physical mass of any other scalar in the spectrum (i.e., playing with the root/pole of its two-point function) it is natural to consider all possible shapes of the scalar spectrum accessible by a given (preferably as small as possible) number of fine-tunings to be at the same footing and let the model parameters just accommodate freely to all the relevant experimental and theoretical constraints. This is the strategy employed many times in the past, e.g., in the non-minimal SU(5) context~\cite{Dorsner:2006hw,Dorsner:2006dj,Bajc:2006ia,Dorsner:2007fy} as well as in the previous analyses~\cite{Bertolini:2012im} of the model under consideration and we shall also stick to this approach here.

\subsubsection{Phenomenology constraints}
\paragraph{Proton lifetime limits.} The recent 90\% C.L. limit on the proton lifetime in the $p\to \pi^0 e^+$ channel reads (see, e.g.,~\cite{Nishino:2012ipa})
\begin{equation}\label{SKlimit}
\tau(p\to \pi^0 e^+) > 8.2\times 10^{33}\,\mathrm{years}.
\end{equation}
In contrast, the Hyper-Kamiokande is assumed to reach the bounds
\begin{align}
\label{HKlimit2025}\tau(p\to \pi^0 e^+)_{{\rm HK},2030} &> 9\times 10^{34}\,\mathrm{years}\,,\\
\label{HKlimit2040}\tau(p\to \pi^0 e^+)_{{\rm HK},2045} &> 2\times 10^{35}\,\mathrm{years}\,,
\end{align}
by years 2030 and 2045, respectively\footnote{These numbers correspond to the sensitivity limits displayed in~\cite{Abe:2011ts} with an extra 5-year offset due to the current delay in the HK timeline.} \cite{Abe:2011ts}. Let us just note that for the kaonic modes the assumed HK limit 
\be
\label{HKlimitKaons2040}
\tau(p\to K^+ \overline{\nu})_{{\rm HK},2045} > 3\times 10^{34}\,\mathrm{years}
\ee 
should be just competitive with the expected LBNE sensitivity reach (assuming the 35kt underground variant); however, in what follows we shall focus only on the pionic mode due to its general preference in non-SUSY GUTs.

\paragraph{Big bang nucleosynthesis constraints.} Any significant extra entropy injected into the primordial plasma during the BBN epoch disturbs the predictions for the abundances of the light elements~\cite{Beringer:1900zz}. Hence, we require there to be no remnants of the high-energy spectrum (in particular, no colored states) with lifetimes longer than a fraction of a second that may be strongly coupled to the plasma.
\subsubsection{Classes of consistent solutions}
Remarkably enough, the initial one-loop analysis~\cite{Bertolini:2012im} revealed that at the single fine-tuning level there are only two classes of solutions conforming all these requirements, namely, those featuring a near-TeV-scale colored octet transforming like $(8,2,+\tfrac{1}{2})$ under the SM and those with an intermediate-scale colored sextet $(6,3,+\tfrac{1}{3})$. In both cases,  the seesaw scale is pushed far above the ESH region and, thus, in turn, the fine-tuning in $\sigma$ is effectively ``traded'' for that in the relevant light scalar mass.

\subsection{TeV-scale octet}
\begin{figure}[t]
\begin{center}
\includegraphics[width=8.5cm]{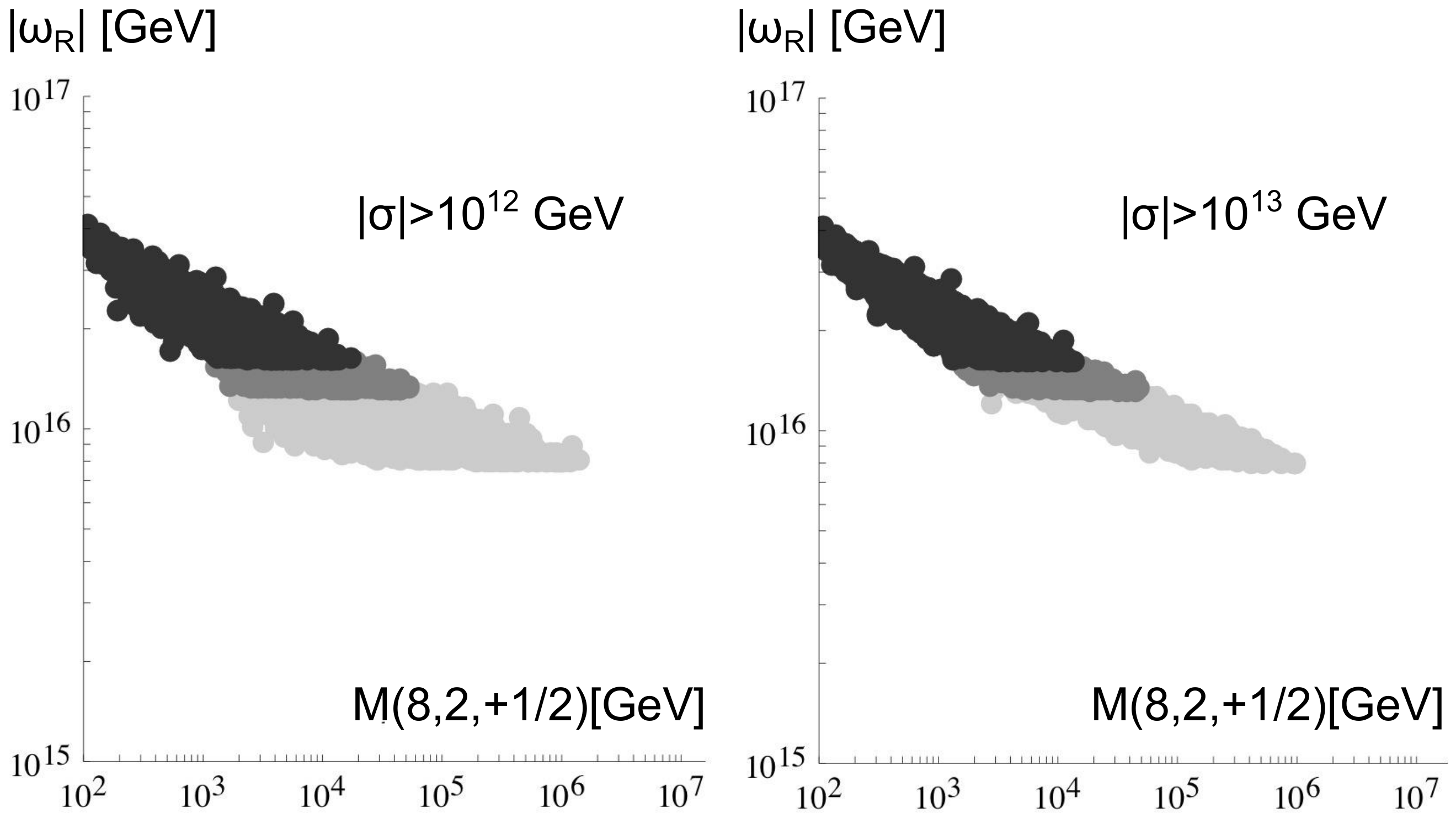}
\end{center}
\caption{Masses of the $(8,2,+\frac{1}{2})$ scalar field allowed by the NLO unification and matter stability constraints (the points consistent with the limits \eqref{SKlimit}, \eqref{HKlimit2025} and \eqref{HKlimit2040} are plotted in light gray, dark gray and black color, respectively), cf.~\cite{Bertolini:2013vta}. In the left panel $|\sigma|\geq 10^{12}\GeV$ is assumed while, in the right panel, we admit for $|\sigma|\geq 10^{13}\GeV$. Remarkably, focusing on the black area, either the octet mass is below $20\,\mathrm{TeV}$ and, thus, potentially, within the reach of the LHC or its near future successors or proton lifetime should be seen at the HK before 2045.  
For more discussion see~\cite{Bertolini:2013vta}.}
\label{FigOctet}
\end{figure}

The former case, i.e., the settings  with the very light  $(8,2,+\tfrac{1}{2})$ have been studied thoroughly in the work~\cite{Bertolini:2013vta}. This dedicated two-loop gauge unification analysis revealed a very interesting tension among the upper limit on the octet mass and the lower bound on the proton lifetime suggesting that either the octet is below about 20 TeV (and, hence, may be within the reach of either LHC or one of its near-future successors) or that proton decay should be seen at the HK, see FIG.~\ref{FigOctet}. Let us remark that this result also illustrates the importance of the NLO gauge unification analysis as, at the LO, the same bounds were so loose (e.g., the upper limit on the octet mass stretched up to about 2000 TeV) that no such phenomenologically interesting feature could have been exploited.  Let us also mention that these limits apply even for the $B-L$  scale as high as $10^{13}\GeV$ which, indeed, is fully compatible with the assumed renormalizable implementation of the seesaw mechanism.

\subsection{ZeV-scale sextet}
As for the second option, namely, the intermediate-scale sextet solutions, only the LO results have been obtained so far, cf. \cite{Bertolini:2012im}. Although this scenario does not seem to provide any striking signal as did the light octet setting, the patches of the parameter space supporting this class of scenarios turned out to be rather small (with the sextet mass $M(6,3,+\tfrac{1}{3})$ stretching from  about $10^{10}$ GeV to about $10^{12}\GeV$), especially for the $B-L$ breaking scale in the seesaw-favored region ($\sigma>10^{12}$~GeV).  Since, as we learned in the octet case, the two-loop effects can change the LO picture considerably, one should check whether the sextet solution is still viable at the NLO level. This  is the scope of the next section.

%%%%%%%%%%%%%%%%%%%%%%%%%%%%%%%%%%%%%%
\section{${\rm \bf ZeV}$-scale sextet at two-loops}
%%%%%%%%%%%%%%%%%%%%%%%%%%%%%%%%%%%%%%
Let us begin with the detailed two-loop gauge unification analysis; the results shall be later on combined with the proton lifetime and other phenomenological constraints in order to assess whether there is still some parameter space left  for the sextet solutions.
\subsection{Two-loop gauge unification\label{sect:running}}
The settings with an intermediate-scale $(6,3,+\tfrac{1}{3})$ are characterized~\cite{Bertolini:2012im} by the generic hierarchy $|\omega_{BL}|\gg |\omega_{R}|$; this, in turn, corresponds to a multi-stage symmetry breaking pattern passing through an intermediate $SU(3)_{c}\otimes SU(2)_{L}\otimes SU(2)_{R}\otimes U(1)_{BL}$ stage at which a number of  components of $\phi$ and $\Sigma$ become massive and can be integrated out. Besides that, there can also be a further step with a yet smaller $SU(3)_{c}\otimes SU(2)_{L}\otimes U(1)_{R}\otimes U(1)_{BL}$ gauge symmetry attained if $|\sigma|$ is larger than $|\omega_{R}|$. At the NLO level, this situation is conveniently modeled by a series of effective gauge theories with the renormalization group (RG) evolution described by their individual $\beta$-functions that, at proper scales, are matched together appropriately, cf.~\cite{Weinberg:1980wa,Hall:1980kf}. For that sake, the details of the (non-tachyonic\!) scalar spectrum are essential, cf. Sect.~\ref{oneloop}.
At the electroweak scale $M_{Z}$ we impose the classical set of boundary conditions
\cite{Beringer:1900zz}
\begin{align*}
  \alpha_s(M_Z)&=0.1185\pm0.0006\,,\\
  \sin^2\theta_W(M_Z)&=0.23126 \pm 0.00005\,,\\
  \alpha_e^{-1}(M_Z)&=127.944\pm0.014\,,
  \end{align*}
where  $\alpha_i\equiv{g_i^2}/{4\pi}$; these numbers are readily translated to
\bea
\label{AlphaErrors}\alpha_3^{-1}(M_{Z})&\equiv&\alpha_s^{-1}(M_{Z})=8.44\pm 0.04,\\
\alpha_2^{-1}(M_{Z})&\equiv&\sin^2\theta_W (M_{Z}) \alpha_e^{-1}(M_{Z})=29.588\pm0.007,\nn\\
\nonumber\alpha_1^{-1}(M_{Z})&\equiv&\frac{3}{5}(1-\sin^2\theta_W(M_{Z})) \alpha_e^{-1}(M_{Z})\nn\\&=&59.013\pm 0.004\nn.
\eea

\subsubsection{Effective gauge theories and matching scales\label{efftheories}}
In what follows there are two basic situations to be distinguished:
\paragraph{$|\sigma|\geq|\omega_R|$.\label{sect:simplerchain}}
In this case there are two symmetry breaking steps in the descend from the $SO(10)$ GUT down to the SM to be characterized by a pair of matching scales $\mu_{1}$ and $\mu_{2}$, namely
\begin{equation}\label{BreakingSimple}
SO(10)\xrightarrow{\mu_2}SU(3)_{c}\otimes SU(2)_{L}\otimes SU(2)_{R}\otimes U(1)_{X} \xrightarrow{\mu_1} \mathrm{SM};
\end{equation}
from now on we shall stick to the canonically normalized version of the $B-L$ charge
\begin{equation}\label{XvsBL}
X=\sqrt{\tfrac{3}{8}}(B-L)\,.
\end{equation}
Numerically, the matching scales $\mu_{1,2}$ will be chosen close to the ``barycenters'' of the sets of fields to be integrated out (typically, in the vicinity of the masses of the gauge bosons associated with the relevant symmetry breaking), in particular
\begin{equation}\label{defmu}
\mu_2 \equiv g|\omega_{BL}|,\quad\mu_1 \equiv g |\sigma|\,;
\end{equation}
in this definition a sample value of the unified coupling at the GUT scale $g=0.56$ was used\footnote{Needless to say, the specific choice of the matching scales is to a large extent irrelevant; the prescription (\ref{defmu}) ensures that the results are only marginally dependent on the specific choices of $\mu_{1,2}$, i.e., that the residual higher-order effects are negligible, cf.~Sect.~\ref{matchingindependence}.}.

\paragraph{$|\sigma|<|\omega_R|$.}
Although we are interested in the solutions with large $|\sigma|$, it may still happen that $|\omega_{R}|$ will be yet bigger. In such a case the relevant symmetry breaking chain can be conveniently extended by a third matching scale $\mu_{1}'$, namely
\bea\label{BreakingComplicated}
SO(10)&\xrightarrow{\mu_2}&SU(3)_{c}\otimes SU(2)_{L}\otimes SU(2)_{R}\otimes U(1)_{X}\nn\\
& \xrightarrow{\mu_1'}& SU(3)_{c}\otimes SU(2)_{L}\otimes U(1)_{R}\otimes U(1)_{X}\nn\\
& \xrightarrow{\mu_1}&\mathrm{SM}
\eea
Numerically, we shall choose
$\mu_1' \equiv g |\omega_R|$;
it is also worth pointing out that, in this case, there are two abelian gauge factors present at the third stage so, in principle, the $U(1)$-mixing effects~\cite{delAguila:1988jz} should be taken into account. Hence, the analysis in this case becomes more involved.

\subsubsection{Two-loop beta functions\label{sect:betas}}
\paragraph{Gauge groups with at most one Abelian factor.}
All the gauge groups in the chain \eqref{BreakingSimple} are of this type, hence, this paragraph fully covers the  $|\sigma|>|\omega_R|$ case. At the two-loop level, the running of the gauge coupling associated with the $i$-th gauge  factor is given by the equation
 \begin{equation}\label{RGE}
  \frac{\mathrm{d}}{\mathrm{d}t}\alpha^{-1}_i=-a_i-\frac{b_{ij}}{4\pi}\alpha_j\,,
  \end{equation}
  where
 $$t=\frac{1}{2\pi}\log\frac{\mu}{M_Z}$$ with $\mu$ corresponding to the running scale.
The coefficients $a_i$ and $b_{ij}$ are computed from the field content of the theory as~\cite{Machacek:1983tz}
\be
\label{acoef}a_i=-\frac{11}{3}C_2(G_i)+\frac{4}{3}\sum_f\kappa_f S_2(F_i) +\frac{1}{3}\sum_s\eta_s S_2(S_i),
\ee
\begin{align}
\label{bcoef}b_{ij}&=\Biggl[-\frac{34}{3}(C_2(G_i))^2\\
\nonumber&+\sum_f\left(4C_2(F_i)+\frac{20}{3}C_2(G_i)\right)\kappa_f S_2(F_i)\\
\nonumber&+\sum_s\left(4C_2(S_i)+\frac{2}{3}C_2(G_i)\right)\eta_s S_2(S_i)\Biggr]\delta_{ij}\\
\nonumber&+4\!\left[\sum_f\kappa_f C_2(F_j) S_2(F_i) +\sum_s\eta_s C_2(S_j) S_2(S_i) \right]\!\!+\!\ldots\,,
\end{align}
where the summations run over all scalar and fermion fields of the theory and $\kappa_f=1$ or $\frac{1}{2}$ for Dirac or Weyl fermions, respectively. Similarly $\eta_s=1$ or $\frac{1}{2}$ for complex or real scalars. Furthermore, $C_2(G_i)$ is the quadratic Casimir operator of the group factor $G_i$, $C_2(F_i)$ and $C_2(S_i)$ are the quadratic Casimirs of the $i$-th group representations $F_i$ and $S_{i}$ and, similarly, $S_2$ are the indexes of the same representation including the multiplicity factors. The ellipsis in the expression \eqref{bcoef} stands for the contributions of the Yukawa couplings which, however,  should have a negligible effect on the running as compared to the gauge interactions (see, e.g., Section IV.D in~\cite{Bertolini:2009qj}).
The system~(\ref{RGE}) has a simple approximate solution
\begin{equation}\label{solalpha}
\alpha_i^{-1}(t)-\alpha_i^{-1}(t_0)=-a_i(t-t_0) +\frac{b_{ij}}{4\pi a_j}\log\left[1-\omega_j(t-t_0)\right]\,,
\end{equation}
where $\omega_j=a_j\alpha_j$, provided $|\omega_j(t-t_0)|\ll1$.
These formulae are relevant for the $SO(10)$ and $SU(3)_{c}\otimes SU(2)_{L}\otimes SU(2)_{R}\otimes U(1)_{X}$ stages of both cases discussed in Sect.~\ref{efftheories} as well as for the ultimate SM running phase. Let us anticipate that a fourth stage may be convenient in the case of the more complicated descent~(\ref{BreakingComplicated}) where the $U(1)$-mixing effects in the $\beta$-functions do play a role, see Sect.~\ref{u1mix}).%
\vskip 2mm
Above $\mu_{2}$, the effective theory is the full $SO(10)$ model with three copies of the 16-dimensional spinor representations accommodating the fermionic matter fields, the 45-dimensional adjoint representation containing the gauge fields,  and the scalar sector consisting of a real 45-dimensional adjoint representation and a complex 126-dimensional (self-dual part of the) 5-index antisymmetric $SO(10)$ tensor. This yields
\be
a=-\frac{37}{3},\qquad b=\frac{9529}{6}.
\ee
\vskip 2mm
At the $SU(3)_{c}\otimes SU(2)_{L}\otimes SU(2)_{R}\otimes U(1)_{X}$ level
all the scalar fields except for the color sextet and the scalars responsible for the subsequent symmetry breaking are integrated out, together with the gauge bosons that became massive at this stage. Hence, the list of survivors comprises the following components (for convenience the fields are classified with respect to the $SU(3)_{c}\otimes SU(2)_{L}\otimes U(1)_{R}\otimes U(1)_{BL}$ quantum numbers, see \eqref{XvsBL}): the gauge bosons residing in the $(8,1,1,0)\oplus(1,3,1,0)\oplus(1,1,3,0)\oplus(1,1,1,0)$ representation, the matter fields living in the three copies of $(3,2,1,+\frac{1}{3})\oplus (\bar{3},1,2,-\frac{1}{3}) \oplus (1,2,1,-1) \oplus(1,1,2,-1)$ and the complex scalars form the $(1,1,3,+2) \oplus(1,2,2,0)\oplus (6,3,1,+\frac{2}{3})$ representation.
There are four RG equations of the type (\ref{RGE}) for $\alpha_{c}$, $\alpha_{L}$, $\alpha_{R}$ and $\alpha_{X}$ with coefficients  given, consecutively, by
  $$
  \def\arraystretch{1.4}
  a=\left(-\frac{9}{2},1,-\frac{7}{3},\frac{13}{2}\right),\qquad
  b=\left(
  \begin{array}{cccc}
  89 & \frac{129}{2} & \frac{9}{2} & \frac{11}{2} \\
  172 & 120 & 3 & \frac{19}{2} \\
  12 & 3 & \frac{80}{3} & \frac{27}{2} \\
  44 & \frac{57}{2} & \frac{81}{2} & \frac{65}{2} \\
  \end{array}
  \right).
  $$
\vskip 2mm
Eventually, at the pure SM level (assuming only one effective Higgs doublet) the $SU(3)_{c}\otimes SU(2)_{L}\otimes U(1)_{Y}$ RG coefficients receive the notorious form~\cite{Machacek:1983tz}
$$
  \def\arraystretch{1.4}
  a=\left(-7,-\frac{19}{6},\frac{41}{10}\right),\qquad
  b=\left(
  \begin{array}{ccc}
  -26 & \frac{9}{2} & \frac{11}{10} \\
  12 & \frac{35}{6} & \frac{9}{10} \\
  \frac{44}{5} & \frac{27}{10} & \frac{199}{50} \\
  \end{array}
  \right).$$

 \paragraph{Multiple Abelian group factors.\label{u1mix}} Finally, let us discuss the fine effects related to the presence of the pair of $U(1)$ factors in the third stage of the chain \eqref{BreakingComplicated}.

Adopting the formalism in which both the $U(1)$ kinetic forms are kept canonical~\cite{delAguila:1988jz} the $U(1)$-mixing effects can be subsumed into an extended form of the covariant derivative including a matrix gauge coupling
\begin{equation}\label{gmatrix}
g\equiv
  \left(\begin{array}{cc}
  g_{RR} & g_{RX}\\
  g_{XR} & g_{XX}
  \end{array}
  \right)\,.
\end{equation}
As explained, for instance, in \cite{Fonseca:2011vn,Fonseca:2013bua}, the Lagrangian of the theory is invariant under the orthogonal field transformations $A_I^\mu\to O_{IJ}A_J^\mu$ where $A^{\mu}$ denotes the $U(1)$ vector boson fields and $O$ is an orthogonal matrix in the relevant field space. Performing, simultaneously, the gauge matrix transformations $g\to gO^T$ the covariant derivative does not change and, hence, the physics remains the same. This redundancy can be removed by considering $gg^T$ instead of $g$; hence, it is very convenient to work with the matrix analogue of the individual $\alpha$ couplings
\begin{equation}\label{defA}
A\equiv \frac{gg^T}{4\pi}.
\end{equation}
As usual, the two-loop RG evolution of the non-abelian couplings depends on the abelian ones, however, in the matrix formalism, their contribution can not be factorized as easily as in Eg.~\eqref{RGE}. This, for $i \in\{c,L\}$, one has instead
\begin{equation}\label{RGEmix2}
\frac{\mathrm{d}}{\mathrm{d}t}\alpha^{-1}_i=-a_i-\frac{b_{ij}}{4\pi}\alpha_j - \frac{c_i}{4\pi}\,,
\end{equation}
where $a_i$ are again the one-loop contributions computed from \eqref{acoef}, $b$ comprises the two-loop contributions from non-abelian couplings only (computed from \eqref{bcoef}; however, since there are only two non-abelian couplings at play, $b$ will be a $2\times 2$ matrix here). Finally, the two-loop contributions of the abelian couplings are calculated as
\begin{align*}
c_i&=4\!\left(\sum_{f,I,J}\kappa_f Q_f^I A_{IJ} Q_f^J S_2(F_i) \!+\! \sum_{s,I,J}\eta_s Q_s^I A_{IJ} Q_s^J S_2(S_i)\!\right).
\end{align*}
where $Q^{I}$ denote the charges of the relevant fields under the $I$-th abelian gauge factor.

The evolution equation for the matrix abelian coupling \eqref{defA} may be written in the form
\begin{equation}\label{RGEmat2}
\frac{\mathrm{d}A^{-1}}{\mathrm{d}t} = - a - \frac{b_i}{4\pi}\alpha_i - \frac{c}{4\pi}
\end{equation}
where (in the current case) $a$, $b_i$ and $c$ are $2\times2$ matrices. As in Eq.~\eqref{RGE}, the matrix $a$ covers the one-loop contributions and, as before, the two-loop contribution was divided into two parts: $b_i$ comprises the contributions from $i$-th non-abelian coupling ($i=c,L$) while $c$ covers the self-interactions in the abelian sector proportional to $A$. The relevant formulae read~\cite{delAguila:1988jz,Bertolini:2009qj}
\bea\label{gamma1}
a_{IJ}&=&\frac{4}{3} \sum_f \kappa_f Q_f^I Q_f^J + \frac{1}{3}\sum_{s}\eta_s Q_s^I Q_{s}^J\,\\
\left(b_k\right)_{IJ}& = &4\left(\sum_{f}  \kappa_f Q_f^I Q_f^J C_2(F_k) + \sum_{s} \eta_s Q_s^I Q_s^J C_2(S_k)
\right)\nn\\
& & + \ldots\\
c_{IJ} &=&4\Biggl(
\sum_{f}  \kappa_f Q_f^I Q_f^J \sum_{K,L}Q_f^K A_{KL} Q_f^L \\
&+& \sum_{s} \eta_s Q_s^I Q_s^J \sum_{K,L}Q_s^K A_{KL} Q_s^L
\Biggr) + \ldots\nn
\eea
where the meaning of all symbols is the same like in Eq.~(\ref{bcoef}) and the parenthesis on $b$ illustrates its structure of a vector of matrices.

After the breaking of the $SU(3)_{c}\otimes SU(2)_{L}\otimes SU(2)_{R}\otimes U(1)_{X}$ %$3_c2_L2_R1_X$ 
symmetry to the %$3_c2_L1_R1_X$
$SU(3)_{c}\otimes SU(2)_{L}\otimes U(1)_{R}\otimes U(1)_{X}$
at the $\mu_1'$ scale the set of the ``light'' fields includes (in the $SU(3)_{c}\otimes SU(2)_{L}\otimes U(1)_{R}\otimes U(1)_{BL}$ notation) the matter fermions in the three copies of $(3,2,0,+\frac{1}{3})\oplus (\bar{3},1,+\frac{1}{2},-\frac{1}{3}) \oplus (\bar{3},1,-\frac{1}{2},-\frac{1}{3}) \oplus (1,2,0,-1) \oplus(1,1,+\frac{1}{2},-1)\oplus(1,1,-\frac{1}{2},-1)$, the complex scalars transforming as $(1,1,-1,+2) \oplus(1,2,+\frac{1}{2},0)\oplus (6,3,0,+\frac{2}{3})$ as well as the relevant vector bosons. With this at hand, the coefficients in \eqref{RGEmix2} can be calculated readily
$$\def\arraystretch{1.4}
(a_c,\,a_L)=(-\tfrac{9}{2},\tfrac{5}{6}),\qquad \left(
\begin{array}{cc}
 b_{cc} & b_{cL} \\
 b_{Lc} & b_{LL} \\
\end{array}
\right)=\left(
\begin{array}{cc}
 89 & \frac{129}{2} \\
 172 & \frac{707}{6} \\
\end{array}
\right)\,,$$
and
\begin{align*}
c_c&=\frac{3}{2}A_{RR}+\frac{13}{6}A_{XX},\\
c_L&=\frac{1}{2}A_{RR}+\frac{17}{6}A_{XX}.
\end{align*}
Furthermore, the elements of the coefficient matrices governing formula \eqref{RGEmat2} read (in the $\{R,X\}$ basis)
$$
 \def\arraystretch{1.4}
a=\left(\begin{array}{cc}
\frac{9}{2}&-\frac{1}{\sqrt{6}}\\
-\frac{1}{\sqrt{6}}&\frac{11}{2}
\end{array}\right),
$$
$$
 \def\arraystretch{1.4}
 b_c=\left(\begin{array}{cc}
12&0\\
0&44
\end{array}\right),
\quad
b_L=\left(\begin{array}{cc}
\frac{3}{2}&0\\
0&\frac{57}{2}
\end{array}\right),
$$
and
\begin{align*}
c_{RR}&=\frac{15}{2} A_{RR} - 4 \sqrt{6} A_{RX} +
\frac{15}{2} A_{XX},\\
c_{RX}&=c_{XR}=-2\sqrt{6} A_{RR} + 15 A_{RX} -
3\sqrt{6} A_{XX},\\
c_{XX}&=\frac{15}{2} A_{RR} - 6 \sqrt{6} A_{RX} +
\frac{29}{2} A_{XX}.
\end{align*}
The resulting set of differential equations \eqref{RGEmix2} and \eqref{RGEmat2} was solved numerically in {\em Mathematica}.

\subsubsection{Threshold corrections}\label{SecThreshold}
A proper matching among the effective gauge theories encompassing the relevant dynamics between the consecutive symmetry breaking scales requires a careful treatment of the threshold corrections~\cite{Weinberg:1980wa,Hall:1980kf}. This, as in the case of the $\beta$-functions, amounts to integrating out the fields that are considered ``heavy'' below the given matching scale.  In the simplest case when a simple gauge group $G$ is spontaneously broken into a direct product of subgroups $G_i$ (with at most one abelian factor) at a certain scale $\mu$, the relevant matching formula reads
\begin{equation}\label{SimpleMatching}
\alpha_i^{-1}(\mu)=\alpha_G^{-1}(\mu) - 4\pi \lambda_i(\mu)
\end{equation}
where (see, for instance,~\cite{Bertolini:2013vta})
\begin{align}\label{FormLambda}
\lambda_i(\mu)&=\frac{1}{48\pi^2}S_2(V_i)+\frac{1}{8\pi^2}\Biggl[-\frac{11}{3}S_2(V_i)\log\frac{M_{V}}{\mu} \\ \nonumber&+\frac{4}{3}\kappa_F S_2(F_i) \log\frac{M_{F}}{\mu}+\frac{1}{3}\eta_S S_2(S_i) \log\frac{M_{S}}{\mu}\Biggr].
\end{align}
Here the arguments $V$, $F$ and $S$ denote the heavy vector bosons, fermions and scalars that are integrated out at the scale $\mu$ and $M_{V}$, $M_{F}$ and $M_{S}$ stand for their masses\footnote{Needless to say, this formula applies only to the case when all the members of the relevant multiplets of $G_{i}$ are degenerate. This implicitly assumes that the subsequent symmetry breaking (that may smear this degeneracy) occurs well below $\mu$ which, however, does not need to be case in general, see the discussion below.}; the rest of notation has been again inherited from \eqref{bcoef}. Let us note that in \eqref{FormLambda} the (Feynman gauge) Goldstone bosons have been included into the scalar part of the expression which, in turn,  makes the formula resemble that for the $a$-coefficient of the one-loop $\beta$-function~(\ref{acoef}); similarly, the FP ghosts have been subsumed into the first factor in the parenthesis. In this form, relation~(\ref{FormLambda}) makes it clear that the effective couplings $\alpha_i^{-1}$ are, at the leading order, independent of the specific choice of $\mu$.

The simple prescription (\ref{SimpleMatching})-(\ref{FormLambda}) makes it relatively straightforward to calculate the threshold corrections for the  $SO(10) \to SU(3)_{c}\otimes SU(2)_{L}\otimes SU(2)_{R}\otimes U(1)_{X}$ breaking at $\mu_2$ which is common to both chains \eqref{BreakingSimple} and \eqref{BreakingComplicated}.
In terms of the $SU(3)_{c}\otimes SU(2)_{L}\otimes SU(2)_{R}\otimes U(1)_{BL}$ quantum numbers the components that decouple at $\mu_2$ are the vector bosons $(3,2,2,-\frac{2}{3})^V\oplus (3,1,1,+\frac{4}{3})^V$ together with the corresponding Goldstones $(3,2,2,-\frac{2}{3})^{GB}\oplus(3,1,1,+\frac{4}{3})^{GB}$ and the scalars $(8,1,1,0)$, $(1,3,1,0)$, $(1,1,3,0)$ and $(1,1,1,0)$ from $\phi$ and $(3,1,1,-\frac{2}{3})$, $(1,3,1,-2)$, $(3,3,1,-\frac{2}{3})$, $(\bar{3},1,3,+\frac{2}{3})$, $(\bar{6},1,3,-\frac{2}{3})$ and $(8,2,2,0)$ from $\Sigma$. 
\vskip 1mm
However, there is a subtlety worth a comment here. As a matter of fact, due to the relative proximity of $\omega_{BL}$ and either $\omega_{R}$ or $\sigma$ and the fine-tuning involved the mass-splittings within the $SU(3)_{c}\otimes SU(2)_{L}\otimes SU(2)_{R}\otimes U(1)_{X}$ multiplets may not be entirely negligible and, hence, the formula~(\ref{FormLambda}) may not be used directly -- note that the spectrum we are working with (see Appendix~\ref{ApSpectrum}) is, indeed, classified with respect to the SM subgroup of the LR symmetry.

Hence, the  classical prescription~(\ref{FormLambda}) should be generalized for such a case. This is facilitated by the fact that the (weighted) index $S_{2}(R_{G})$ of a representation $R_{G}$ calculated from the generators of a larger group $G$ is easily decomposed into the sum of the (weighted) indexes $S_{2}(R_{H}^{i})$ of the components $R_{H}^{i}$  of $R_{G}$ decomposed under its subgroup $H$. In this case, all structures in~(\ref{FormLambda}) of the type $S_{2}(R_{G})\log (M_{R_{G}}/\mu)$ may be just replaced by
\be\label{FormLambda2}
\sum_{i}S_{2}(R_{H}^{i})\log (M_{R_{H}^{i}}/\mu)\,,
\ee
which, in the exact degeneracy limit reduces to the previous form.

Nevertheless, this approach is not entirely straightforward as one also has to take into account that, in principle, there may be significant thresholds from ``off-diagonal'' vacuum polarization graphs if the $SU(2)_{R}$ is broken close to the $SO(10)$ scale and, accidentally, the different components of some of the $SU(2)_{R}$ multiplets happen to be significantly spread in masses (in comparison to the ``reference'' scale $\omega_{BL}$). In such a case, the ``$R-X$-mixing'' graphs do not drop (as it would be obviously the case in the degenerate limit due to the zero trace of the $SU(2)_{R}$ Cartan). Hence, technically, one should either retain the light members of the $SU(2)_{R}$ multiplets throughout the effective LR stage and integrate them out only at $\mu_{1}'$ or, alternatively, carry on the information about the sizable off-diagonal $R-X$ thresholds down to the subsequent matching scale (again, $\mu_{1}'$). These approaches are technically equivalent (up to tiny higher order effects) for the heavy scalars; however, for the vectors, the former is not an option as the formalism introduced in Sect.~\ref{sect:betas} (in particular, formula (\ref{bcoef})) is suitable only for vectors in the adjoint representation of the relevant gauge group. Therefore, we shall adopt the latter strategy of integrating out the entire $SU(2)_{R}$ multiplets of the relevant vectors and scalars (see the list below) at $\mu_2$. Hence, besides the ``standard'' non-abelian threshold functions $\lambda_{c}(\mu_2)$ and  $\lambda_{L}(\mu_2)$
we introduce a ``threshold matrix''
\begin{equation}\label{matLambda}
\Lambda=\left(\def\arraystretch{1.4}\begin{array}{cc} \lambda_{RR}&\lambda_{RX}\\ \lambda_{XR}& \lambda_{XX}\end{array}\right)\,,
\end{equation}
that will keep track of the off-diagonal ($\mu$-independent) non-degeneracy effects until the subsequent $\mu_{1}$-scale matching where these will be ``collapsed'' appropriately into the SM hypercharge factor $\lambda_{Y}$. The entries of $\Lambda$ are given by the general formula
\begin{align}\label{lambdaMat}
\Lambda_{IJ}(\mu)&=\frac{1}{48\pi^2}Q_V^I Q_V^J + \frac{1}{8\pi^2}\Biggl[-\frac{11}{3}Q_V^I Q_V^J\log\frac{M_{V}}{\mu} \\
\nonumber&+ \frac{4}{3} \kappa_F Q_F^I Q_F^J  \log\frac{M_{F}}{\mu}+ \frac{1}{3}\eta_S Q_S^I Q_{S}^J \log\frac{M_{S}}{\mu}\Biggr].
\end{align}
where $I$ and $J$ run over $R$ and $X$ and $Q^{I,J}$ denote the relevant Cartans at play, i.e., the generators of the $U(1)_{R}\otimes U(1)_{X}$ subgroup of the $SU(2)_{R}\otimes U(1)_{X}$ gauge symmetry.

Given this, the resulting matching conditions at $\mu_{2}$ read
\begin{align}
\label{SU3ccondatmu2}\alpha_c^{-1}(\mu_2)&=\alpha_G^{-1}(\mu_2) - 4\pi \lambda_c(\mu_2)\,,\\
\label{SU2Lcondatmu2}\alpha_L^{-1}(\mu_2)&=\alpha_G^{-1}(\mu_2) - 4\pi \lambda_L(\mu_2)\,,\\
\label{trickycondition}A^{-1}(\mu_2)&=\alpha_G^{-1}(\mu_2)\unit - 4\pi \Lambda(\mu_2)\,,
\end{align}
where the diagonal entries of the $A$ matrix encode the initial conditions for the $SU(2)_{R}\otimes U(1)_{X}$ couplings $\alpha_{R}$ and $\alpha_{X}$ at $\mu_{2}$, respectively,  while its off-diagonalities serve as the book-keeping of the aforementioned heavy-field non-degeneracy effects and, thus, are not subject to any RG evolution  throughout the  $SU(3)_{c}\otimes SU(2)_{L}\otimes SU(2)_{R}\otimes U(1)_{X}$stage. The explicit form of all the  $\lambda$ factors in formulae (\ref{trickycondition}) is written in Appendix~\ref{app:mu2}. 

Let us also note that employing such a ``matrix'' notation already at this level is very convenient even if there is no genuine effective $SU(3)_{c}\otimes SU(2)_{L}\otimes U(1)_{R}\otimes U(1)_{X}$ stage to be considered (as, e.g., in the $|\sigma|>|\omega_{R}|$ case) because it simplifies the subsequent hypercharge matching -- rather than two different prescriptions there will be a single matching formula valid for both VEV hierarchies discussed in Sect.~\ref{sect:simplerchain}.
\vskip 1mm

Next, let us discuss the matching among the effective LR stage and the genuine $SU(3)_{c}\otimes SU(2)_{L}\otimes U(1)_{R}\otimes U(1)_{X}$-symmetric effective gauge theory at the $\mu_{1}'$ scale if the symmetry breaking chain (\ref{BreakingComplicated}) is invoked, cf.~Sect.~\ref{u1mix}. Here, the effective theory does feature the dynamical $U(1)$-mixing effects and, hence, it is mandatory\footnote{Barring the alternative scheme with the $U(1)$ couplings kept diagonal all the time and working with a non-canonical kinetic form in the abelian sector, cf.~\cite{Luo:2002iq}.} to use the matrix arrangement of the gauge couplings in the abelian sector, cf. Sect.~\ref{u1mix}, as well as the matrix form of the corresponding thresholds~(\ref{matLambda}).
The fields that are integrated out at  $\mu_1'$ are, namely, the vector bosons $(1,1,\pm 1,0)^V$, the corresponding Goldstone boson $(1,1,0,+2)^{GB}$ and the scalars $(1,1,+1,+2)$, $(1,2,-\frac{1}{2},0)$. Let us note that the last scalar comes (together with the SM Higgs left doublet) from the bi-doublet $(1,2,2,0)$ and we adjust its mass to the $SU(2)_R$ breaking scale by hands.
In practice, this corresponds to working with an admixed extra 10-dimensional scalar representation decoupled at $M_{G}$ and, thus, mimicking the setting with a potentially realistic Yukawa sector. For further comments on this issue the reader is deferred to Sect.~\ref{Par10H}.

As anticipated, thanks to the matrix form of the initial condition~(\ref{trickycondition}), the  matching formulae at $\mu_{1}'$ are  simple 
\begin{align}
\nonumber\alpha_c^{-1}(\mu_{1-}')&=\alpha_c^{-1}(\mu_{1+}') - 4\pi \lambda_c(\mu_1')\,,\\
\label{mu1primematching}\alpha_L^{-1}(\mu_{1-}')&=\alpha_L^{-1}(\mu_{1+}') - 4\pi \lambda_L(\mu_1')\,,\\
\nonumber A^{-1}(\mu_{1-}')&=A^{-1}(\mu_{1+}') - 4\pi \Lambda(\mu_1')\,.
\end{align}
As usual, we dare to use the same symbols for the running couplings in the $SU(3)_{c}\otimes SU(2)_{L}\otimes SU(2)_{R}\otimes U(1)_{X}$ stage on the LHS and those of the $SU(3)_{c}\otimes SU(2)_{L}\otimes SU(2)_{R}\otimes U(1)_{X}$ on the RHS of Eqs.~(\ref{mu1primematching}).
The threshold factors above are given in Appendix~\ref{app:matchingmu1prime}.
\vskip 1mm

Finally, let us consider the matching of either the $SU(3)_{c}\otimes SU(2)_{L}\otimes SU(2)_{R}\otimes U(1)_{X}$ or the $SU(3)_{c}\otimes SU(2)_{L}\otimes U(1)_{R}\otimes U(1)_{X}$  effective theory to the SM at the  $\mu_1$ scale.
As for the former, the fields to be integrated out at $\mu_{1}$ are (in the SM notation): the vector bosons $(1,1,0)^V\oplus(1,1,\pm 1)^V$ together with the associated Goldstones $(1,1,0)^{GB}\oplus(1,1,+1)^{GB}$, the singlet real scalar $(1,1,0)$ and the complex scalars $(1,1,+2)$ and $(1,2,-\frac{1}{2})$. In the latter case, the basic set of ``heavy'' fields here is almost trivial as there are only full-signet vector bosons and scalars there. In both cases, we also integrate out the light sextet $(6,3,+\frac{1}{3})$ at $\mu_{1}$.

Technically, the threshold factors $\lambda_{c, L,Y}$ may be again obtained right from the formula~\eqref{FormLambda}; the hypercharge matching, however, is more complicated due to the rank reduction. Using the well-known relation for the (canonically normalized) SM hypercharge $Y=\sqrt{\frac{3}{5}} T_R^3 + \sqrt{\frac{2}{5}} X$ the relevant matching formulae read
\begin{align}
\nonumber\alpha_c^{-1}(\mu_{1-})&=\alpha_c^{-1}(\mu_{1+}) - 4\pi \lambda_c(\mu_1),\\
\label{MatchMu1}\alpha_L^{-1}(\mu_{1-})&=\alpha_L^{-1}(\mu_{1+}) - 4\pi \lambda_L(\mu_1),\\
\nonumber\alpha_Y^{-1}(\mu_{1-})&= P_{Y} A^{-1}(\mu_{1+})P_{Y}^T - 4\pi \lambda_Y(\mu_1)
\end{align}
where $P_{Y}=\left(\sqrt{\frac{3}{5}},\sqrt{\frac{2}{5}}\right)$ is the (first row of the) corresponding ``hypercharge projector''. As before, we overload the notation for the non-abelian running couplings; the explicit form of the threshold functions\footnote{These, however,  differ in the $SU(3)_{c}\otimes SU(2)_{L}\otimes SU(2)_{R}\otimes U(1)_{X}$ and $SU(3)_{c}\otimes SU(2)_{L}\otimes U(1)_{R}\otimes U(1)_{X}$ cases due to the ``richer'' set of dynamical fields necessary to break the former symmetry straight to the SM and a higher number of the associated Goldstone bosons/massive vectors in the former case.} is given in Appendix~\ref{app:mu1matchingA} and \ref{app:mu1matchingB}.

%%%%%%%%%%%%%%%%%%%%%%%%%%%%%%%%%%
\subsection{Proton Decay\label{d6protondecay}}
%%%%%%%%%%%%%%%%%%%%%%%%%%%%%%%%%%
\paragraph{Gauge induced $d=6$ proton decay.} Given the generic preference of the pion decay modes in non-SUSY GUTs which, at the same time, are in the focus of many of the existing and future experiments, in what follows we shall concentrate entirely on the $p\to \pi^0 e^+$ decay channel. Assuming no extra flavor suppression in the relevant baryon-number-violating currents\footnote{Needless to say, without a detailed analysis of the flavor structure of the model under consideration this may be seen as a strong assumption; however, it corresponds to the rather natural expectation of no accidental cancellation in the product of the unitary matrices parametrizing the charged BNV currents which, in turn, should have the relevant entries in the ${\cal O}(1)$ ballpark.} the corresponding partial decay width is, in the $SO(10)$ context, given by~\cite{Nath:2006ut}
\begin{align}\label{Gamma}
&\Gamma(p\to \pi^0 e^+)=\frac{\pi\, m_p\,\alpha _G^2 }{4
   f_{\pi}^2}  |\alpha|^2 A_L^2 (D+F+1)^2 \\
  \nonumber &\times\left(A_{SR}^2 \left(\frac{1}{M_{(X',Y')}^2}+\frac{1}{M_{(X,Y)}^2}\right)^2 + \frac{4 A_{SL}^2}{M_{(X,Y)}^4}\right)\,,
\end{align}
where $m_p$ is the proton mass, $M_{(X,Y)}$ and $M_{(X',Y')}$ are the masses of the heavy vector bosons with the SM quantum numbers $(3,2,-5/6)$ and $(3,2,+1/6)$, and $\alpha_G$ is the gauge coupling at the unification scale. Furthermore, $f_\pi=139\,\mathrm{MeV}$, $\alpha=0.009\GeV^3$ and $D+F=1.267$ are the phenomenological factors obtained in the chiral perturbation theory and lattice studies (their specific values were taken from the reference~\cite{Nath:2006ut}).
The one-loop evolution of the effective four-fermion BNV operators~\cite{Buras:1977yy,Ellis:1979hy} in the low-energy domain (i.e., from the proton mass to the electroweak scale~\cite{Nath:2006ut}) is taken care of  by the coefficient $A_L\approx1.4$  while the $A_{SL(SR)}$ factors  
\begin{equation}\label{ASLR}
A_{SL(SR)}=\prod_{i=1}^3\prod_{x}^{m_Z\leq m_x<M_G}\left[\frac{\alpha_i(m_{x+1})}{\alpha_i(m_x)}\right]^{\frac{\gamma_{L(R)}^{i}}{\sum_y^{M_Z\leq M_y<m_P}\Delta a^i_y}}
\end{equation}
contain the running effects from $M_{Z}$ to $M_{G}$. The relevant anomalous dimensions read
$\gamma_L=(\frac{23}{20},\frac{9}{4},2)$ and $\gamma_R=(\frac{11}{20},\frac{9}{4},2)$; the symbols $x$ and $y$ label the fields driving the RG evolution at each stage and $\Delta a_y^i$ is the contribution of the field $y$ to the one-loop beta function for the $i$-th coupling. In what follows we shall compare the width~(\ref{Gamma}) with the existing SK proton lifetime limit (\ref{SKlimit}) as well as with the expected HK sensitivity bounds (\ref{HKlimit2025}) and (\ref{HKlimit2040}).

\paragraph{Scalar induced $d=6$ proton decay.} Since the scalar-driven $d=6$ amplitudes are often suppressed by the smallness of the first-generation Yukawa couplings, the bounds on the mass of the ``dangerous'' mediators are less strict than the bounds on the gauge bosons. We require the mass of the scalar leptoquark with the SM quantum numbers $(3,1,-\frac{2}{3})$ to exceed the (rather conservative) bound of $10^{14}\GeV$; this, however, does not restrict the allowed parameter space at all as this mass always turns out to be near the GUT scale.

\paragraph{d=7 proton decay.} Although the $d=7$ operators are usually highly suppressed due to the extra inverse powers of the mediator mass, one has to be careful when some of the fields are pulled down far below the GUT scale (which is exactly the case for the $(6,3,+\frac{1}{3})$ scalar in our analysis). However, going through the potentially dangerous $d=7$ operators listed, e.g., in \cite{Babu:2012vb}, one finds that this field does not participate in  such interactions.

%%%%%%%%%%%%%%%%%%%%%%%%%%%%%%
\subsection{Absolute neutrino mass scale\label{seesawscale}}
%%%%%%%%%%%%%%%%%%%%%%%%%%%%%%
There is one more assumption worth a comment that we shall make in what follows; in particular, we shall impose a lower bound on the size of the rank-breaking VEV $\sigma$. Since this parameter, together with the Yukawa coupling of $\Sigma$, governs the mass scale of the RH neutrinos, the seesaw-generated light neutrino masses are inverse proportional to $\sigma$. Assuming no accidental cancellation in the Dirac neutrino mass matrix (thus adopting  the minimal fine-tuning policy advocated in Sect.~\ref{minimalfinetuning} in the Yukawa sector of the model) the $B-L$ breaking VEV $\sigma$ should fall into the $10^{12-14}$ GeV ballpark; let us note that this region is also indicated by the existing Yukawa fits, cf.~\cite{Joshipura:2011nn,Dueck:2013gca}.  From now on, we shall mostly stick to this ``natural'' domain for $\sigma$; for further comments an interested reader is deferred to Sect.~\ref{conclusionsoutlook}.

%%%%%%%%%%%%%%%%%%%%%%%%%%%%%%%%%
\subsection{Results}
%%%%%%%%%%%%%%%%%%%%%%%%%%%%%%%%%
Let us start with the basic description of the regions of the parameter space that turn out to be consistent with all the ``hard'' constraints discussed in Sect.~\ref{sect:consistency}, namely, the perturbativity, unification, proton lifetime etc. Later on, we shall comment on the important role a possible lower limit on the seesaw scale may play in a further reduction of the allowed domain.  

\subsubsection{The NLO gauge unification and proton decay constraints}
\begin{figure}
\begin{center}
\includegraphics[width=8cm]{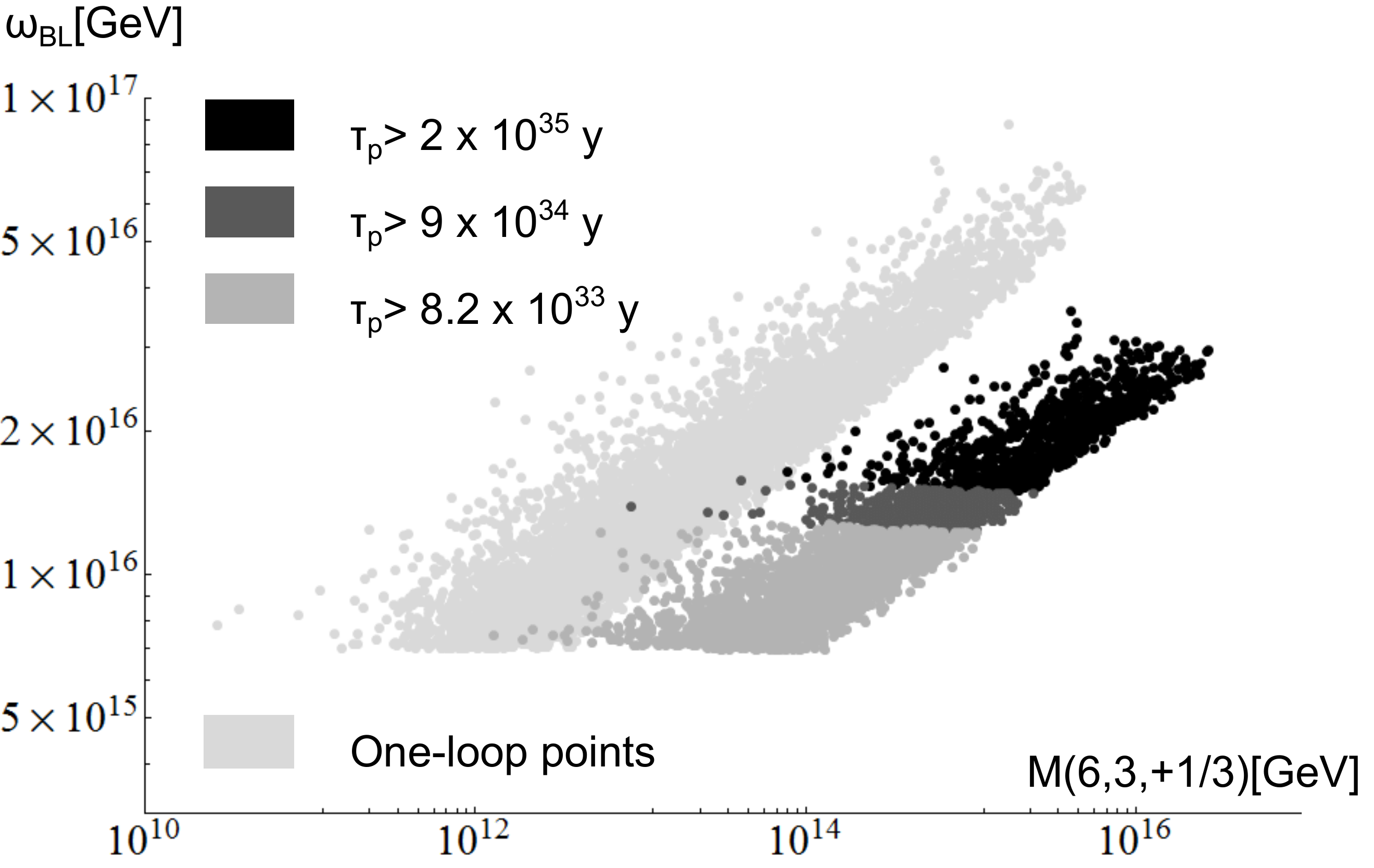}
\end{center}
\caption{The points consistent with two-loop unification constraints and the limits \eqref{SKlimit}, \eqref{HKlimit2025} and \eqref{HKlimit2040} on the proton lifetime plotted in light gray, dark gray and black color, respectively. The light-gray band in the background encloses the settings identified in the previous one-loop  analysis~\cite{Bertolini:2012im} that are consistent with the current SK proton lifetime limit \eqref{SKlimit}.}
\label{FigM6BL}
\end{figure}

The shape of the parameter space that consistently supports the intermediate-scale sextet solutions at the NLO level is  similar to that identified in the one-loop analysis~\cite{Bertolini:2012im}, namely $\omega_{BL}>0$, $\beta_4'<0$,  $\beta_4>0$, $a_0>-0.1$ and $|\gamma|<0.6$; on the other hand, all the dimensionful parameters were shifted considerably. First, while the mass of the sextet  $M(6,3,+\frac{1}{3})$ was increased by factor of about $30$, the maximum NLO-allowed $\omega_{BL}$ was lowered by a factor of $2.5$ which, in turn, reduced considerably the volume of the parameter space consistent with the considered proton lifetime limits. In FIG.~\ref{FigM6BL} the points consistent with the two-loop unification are plotted in three different shades of gray distinguishing among those consistent with the three proton decay bounds \eqref{SKlimit}, \eqref{HKlimit2025}, and \eqref{HKlimit2040} (black points correspond to the strongest limit). For comparison, in the same plot, the points consistent with the current SK proton decay limit \eqref{SKlimit}  at the LO level are shown in light gray\footnote{Let us note that the shape of the allowed one-loop parameter space depicted in FIG.~\ref{FigM6BL} is different from that given in~\cite{Bertolini:2012im} where, for technical simplicity, only points for which the sextet was the lightest of the ``heavy'' fields were considered.}.

\begin{figure}
\begin{center}
\includegraphics[width=8cm]{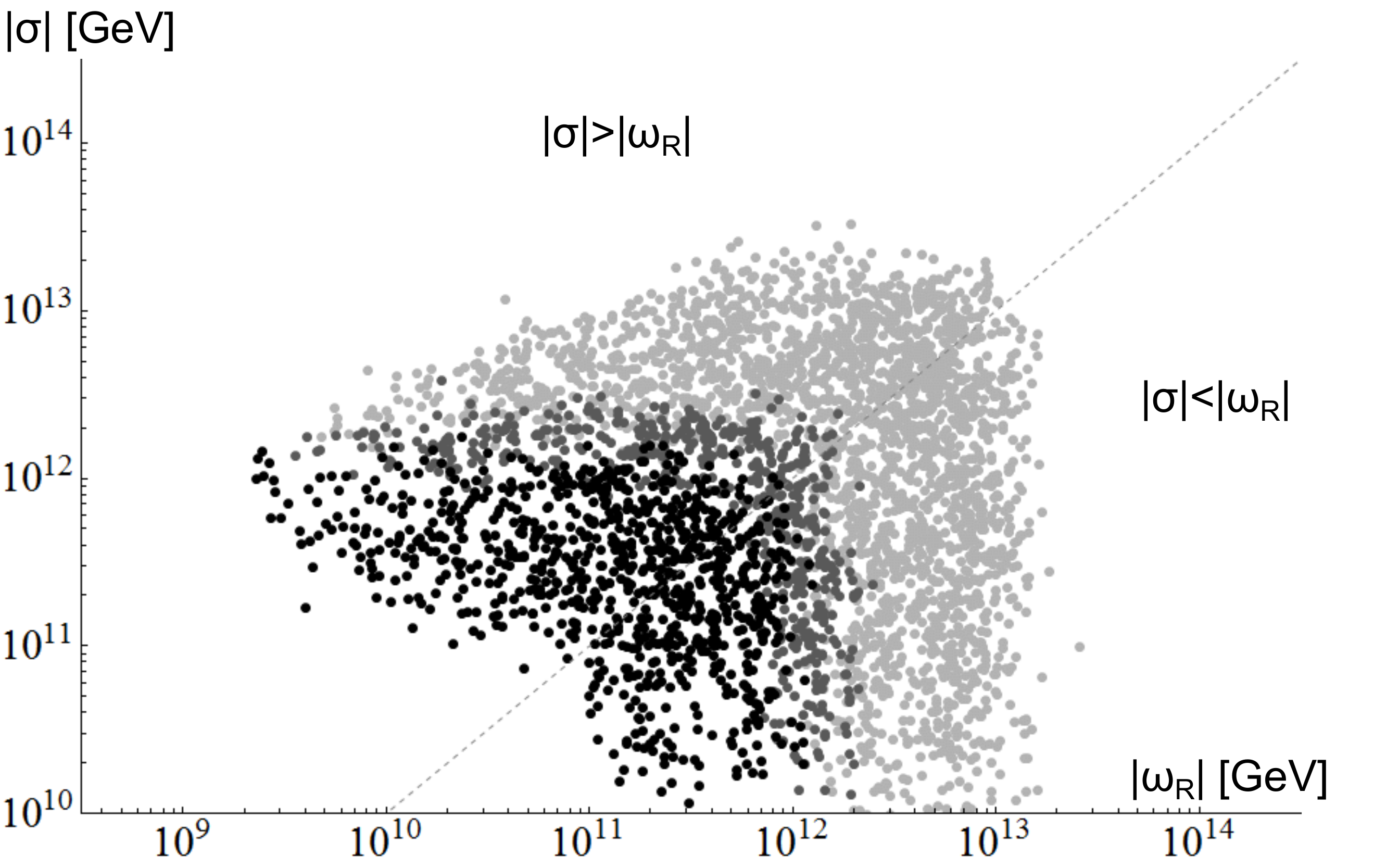}
\end{center}
\caption{The values of $|\omega_R|$ and $|\sigma|$ for the points fulfilling the unification and proton lifetime constraints at the NLO level (color code as in FIG.~\ref{FigM6BL}). The dashed line corresponds to $|\omega_R|=|\sigma|$; above and below this line different symmetry breaking chains have been implemented, cf. Eqs.~\eqref{BreakingSimple}, \eqref{BreakingComplicated}.}
\label{FigSigmaOmegaR}
\end{figure}

The allowed parameter space in the $|\omega_R|$-$|\sigma|$ projection is depicted in FIG.~\ref{FigSigmaOmegaR}; the two qualitatively different regions above and below the diagonal line correspond to the two different symmetry breaking chains considered in Sect.~\ref{efftheories}. 
It is clear that, indeed, $\mathrm{max}\{|\omega_R|,|\sigma|\} \ll \omega_{BL}$ which justifies the selection of the effective $SU(3)_{c}\otimes SU(2)_{L}\otimes SU(2)_{R}\otimes U(1)_{X}$ stage in Sect.~\ref{efftheories}. Moreover, both the regimes with either $|\omega_R|>|\sigma|$ or $|\omega_R|< |\sigma|$ do occur among the consistent points indicating that the bifurcation of the subsequent part of the symmetry breaking chain \eqref{BreakingSimple} and \eqref{BreakingComplicated} is meaningful. Let us note that the parameter space extends rather far from the $|\sigma|=|\omega_R|$ diagonal since, in both cases, the lower VEV can affect the running only marginally; in fact, for $|\sigma|<|\omega_R|$ the fields associated to the symmetry breaking at the $\mu_1$ scale are even full SM singlets and, as such, they leave the beta-function of the ``effective hypercharge'' intact; hence, the ``width'' of the region under the diagonal line does not depend on $\mu_{1}$ (and, hence, neither on $\sigma$). Thus, for $|\sigma|<|\omega_R|$, there is in principle no lower limit on $\sigma$ from the gauge unification constraints, cf.~\cite{Bertolini:2009qj}.

\subsubsection{Seesaw scale constraints}
However, as anticipated in Sect.~\ref{seesawscale},  the settings with very small $\sigma$'s suffer from the issues with the absolute neutrino mass scale unless the Dirac neutrino mass matrix is made artificially small. In the rest of this section we shall adopt extra the constraint $|\sigma|\geq 10^{12}\GeV$ and illustrate its enormous discriminative power.

\begin{figure}[t]
\begin{center}
\includegraphics[width=8cm]{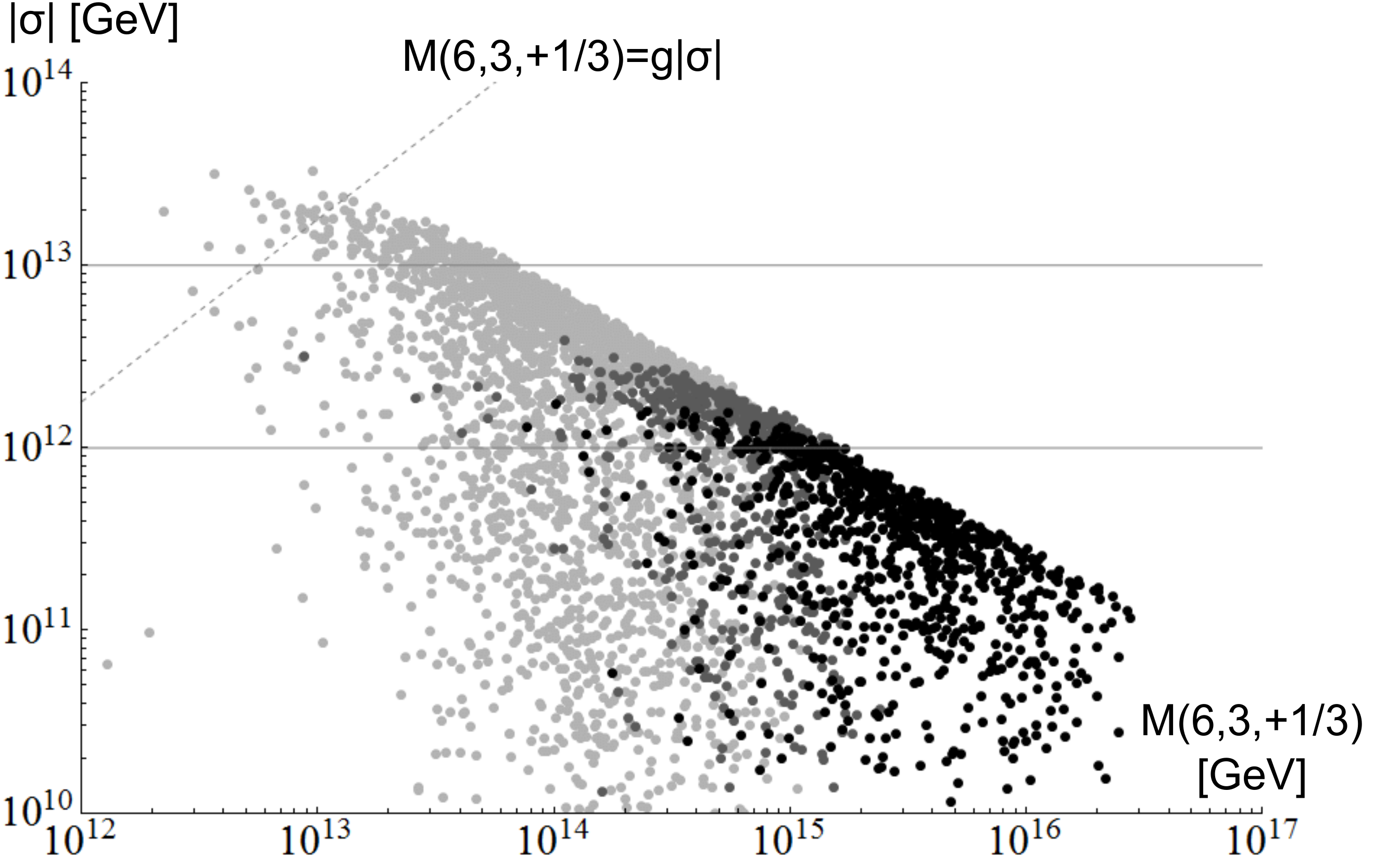}
\end{center}
\caption{$|\sigma|$ as a function of $M(6,3,+\frac{1}{3})$ for the NLO solutions consistent with all the requirements of Sect.~\ref{sect:consistency} (color code as in FIG.~\ref{FigM6BL}). The sharp boundary in the N-E  direction reflects the unification constraints in the case of $|\sigma|>|\omega_R|$, on the other hand, the settings with $|\omega_R|>|\sigma|$ stretch far from this edge since the value of $|\sigma|$ does not affect the unification pattern in such cases. The solid lines correspond to $|\sigma|=10^{12}\GeV$ and $|\sigma|=10^{13}\GeV$ levels, respectively. Recall that $\sigma$ governs the  seesaw scale as well as the amount of fine-tuning necessary to obtain a realistic light neutrino mass spectrum and mixing. }
\label{FigM6Sigma}
\end{figure}

\begin{figure}[h]
\begin{center}
\includegraphics[width=8.5cm]{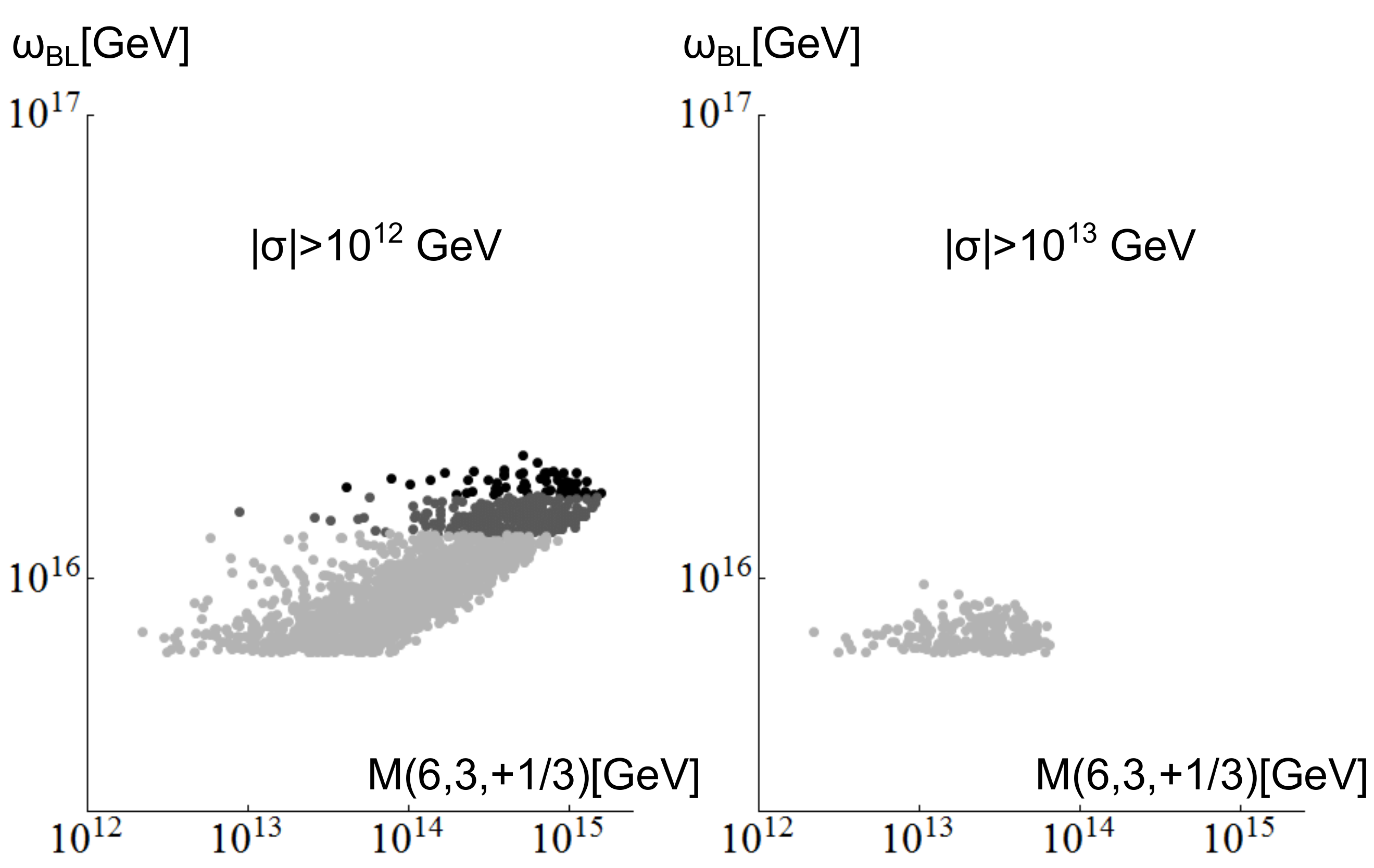}
\end{center}
\caption{The same as in FIG.~\ref{FigM6BL} with an extra assumption of $\sigma>10^{12}\GeV$ (on the left) and $\sigma>10^{13}\GeV$ (on the right). Notice that in the latter case the entire allowed parameter space may be probed by Hyper-K by 2030, see Eq.~(\ref{HKlimit2025}).}
\label{FigM6BLShrink}
\end{figure}

To this end, let us begin with FIG.~\ref{FigM6Sigma} which shows that the consistent values of $|\sigma|$ decrease with growing $M(6,3,+\frac{1}{3})$. Recalling that $\omega_{BL}$ and, hence, the proton lifetime also grow along this direction it is not surprising that once $|\sigma|\geq 10^{12}\GeV$ is required only few points consistent with the 2045 HK limit survive. Remarkably enough, {\em for $|\sigma|\geq 10^{13}\GeV$,  the whole consistent domain is covered by the Hyper-K sensitivity band.} This behavior is best seen in FIG.~\ref{FigM6BLShrink} where the parameter space from FIG.~\ref{FigM6BL} is further constrained by the requirements  of $|\sigma|\geq 10^{12}\GeV$ and $|\sigma|\geq 10^{13}\GeV$, respectively.


\subsubsection{Examples}
\begin{figure}[t]
\begin{center}
\includegraphics[width=8cm]{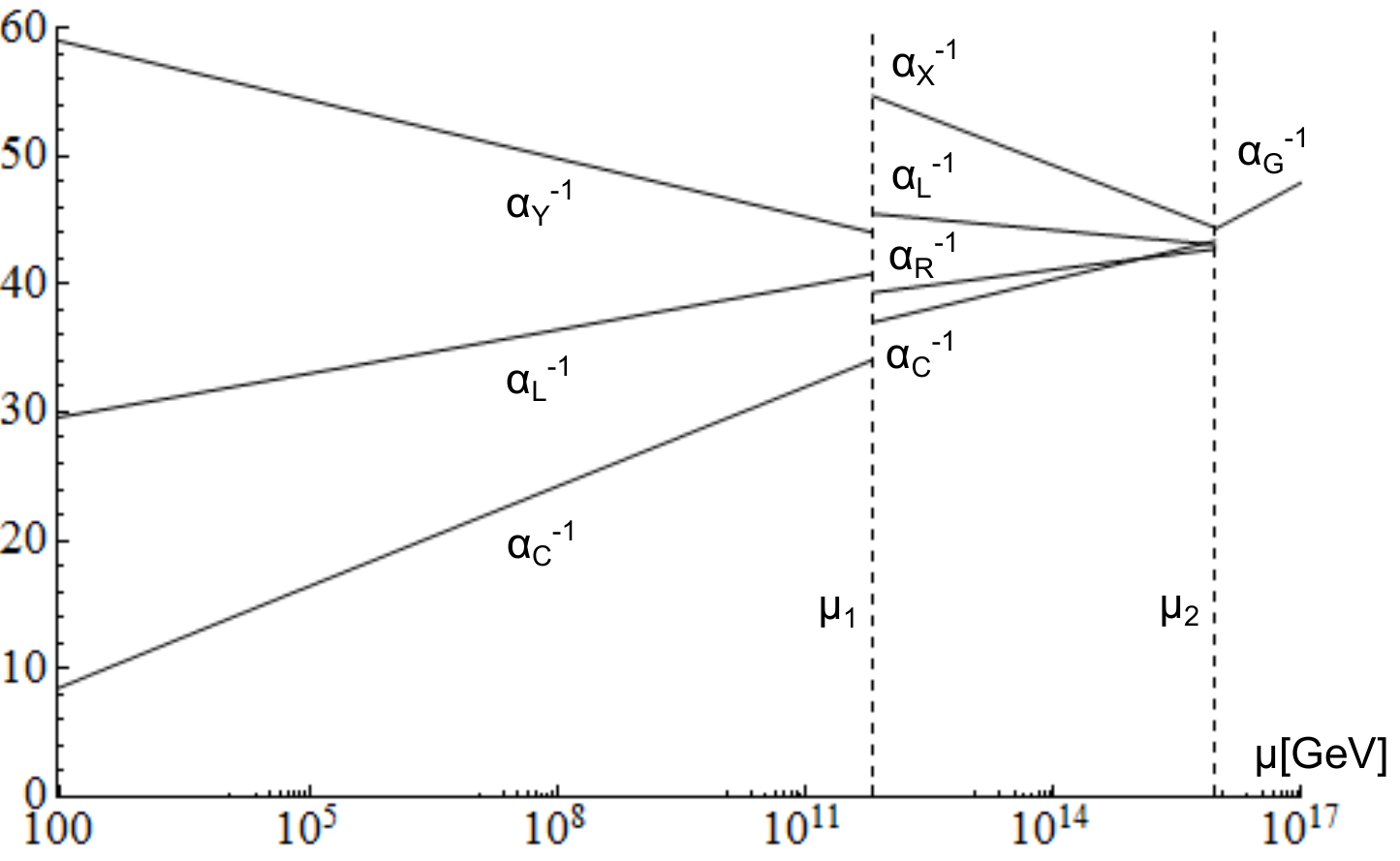}
\end{center}
\caption{The gauge unification pattern for the sample Point 1 in TABLE~\ref{TabPar}; the RG evolution passes through the $SO(10)$, $3_c2_L2_R1_X$ and SM stages, respectively.}
\label{FigRun1}
\end{figure}

\begin{table}
\def\arraystretch{1.2}
\begin{tabular}{| >{$}c<{$} | >{$}c<{$} | >{$}c<{$} |}
\hline
 & \text{Point 1} & \text{Point 2}\\
\hline
 \omega_R & -5.1\times 10^{10}\GeV & -4.1\times 10^{12}\GeV \\
 \omega_{BL}& 1.6\times 10^{16}\GeV & 8.8\times 10^{15}\GeV \\
 \sigma & 1.2\times 10^{12}\GeV & 1.1\times 10^{13}\GeV \\
 \tau & -2.0\times 10^{16}\GeV & -3.2\times 10^{15} \GeV\\
a_0& 0.07 & 0.92 \\
\alpha& 0.93 & 0.14 \\
\beta_4& 0.98 & 0.68 \\
\beta'_4& -0.34 & -0.09 \\
\gamma_2& 0.30 & 0.23 \\
\lambda_0& 0.76 & -0.19 \\
\lambda_2& 0.51 & -0.68 \\
\lambda_4& -0.10 & 0.88 \\
\lambda_4'& 0.56 & -0.37 \\ \hline
M(6,3,+\frac{1}{3})& 1.0\times 10^{15} \GeV& 2.2\times 10^{13} \GeV\\
\hline
\tau_p& 3\times 10^{35}\, \mathrm{y}& 2\times 10^{34}\, \mathrm{y}\\
\hline
\end{tabular}
\caption{A pair of sample points consistent, simultaneously, with the NLO unification constraints, the SK proton lifetime limit~(\ref{SKlimit}) and with the extra assumption $|\sigma|>10^{12}\GeV$. Let us note that Point 1 would satisfy also the expected 2045 HK limits \eqref{HKlimit2040} (but it needs  $|\sigma|<10^{13}\GeV$) while Point~2 obeys the $|\sigma|>10^{13}\GeV$ constraint (but the associated proton decay signal would then be revealed at the HK), but none of them satisfies both these bounds.}
\label{TabPar}
\end{table}
In TABLE \ref{TabPar} we show two examples of the consistent  settings where the aforementioned correlations between the dimensionful parameters can be seen explicitly. In the first case (Point 1 in  TABLE~\ref{TabPar}), we have chosen one of the black points in the left-hand part of FIG.~\ref{FigM6BLShrink} for which the estimated proton lifetime reaches up to $3\times 10^{35}$~years; as expected, the value of $|\sigma|=1.2\times 10^{12}\GeV$ turns out to be rather low. On the other hand, the mass of the sextet is just slightly fine-tuned from its natural position at around $M_{G}$: $M(6,3,+\frac{1}{3})\approx 10^{15}\GeV.$\footnote{This, indeed, agrees with~\cite{Bertolini:2009qj} where it was shown that the model with the GUT-scale breaking driven by $45_H$ suffers from a very low seesaw scale if no extra fine-tuning is invoked.} The hierarchy of the relevant VEVs  corresponds to the breaking chain~\eqref{BreakingSimple} and, thus, there are two matching scales depicted in FIG.~\ref{FigRun1}. Let us note that the high mass of the sextet here leads to sizable threshold corrections to non-abelian couplings at $\mu_1$, cf. Sect.~\ref{parM6}.

For Point~2 in TABLE~\ref{TabPar} the proton lifetime $\tau\approx 2\times 10^{34}$~years is just above the current SK limit, however, the seesaw scale $\sigma$ exceeds $10^{13}\GeV$ and, hence, gives rise to a very comfortable setting for neutrinos. However, such a ``large'' $\sigma$ can be achieved only for the price of a rather light sextet: $M(6,3,+\frac{1}{3})\approx2.2\times 10^{13}\GeV$. This point also corresponds to the symmetry breaking chain~\eqref{BreakingSimple} but, in comparison with the former case, the $3_c2_L2_R1_X$ symmetry stage is short and the threshold corrections to $\alpha_L^{-1}$ and $\alpha_c^{-1}$ at $\mu_1$ are much smaller because the sextet is rather close to this matching scale (see FIG.~\ref{FigRun2}).
For both cases, the shape of the ``heavy'' spectrum is detailed in Appendix~\ref{ApSpectrum}.
\begin{figure}[t]
\begin{center}
\includegraphics[width=8cm]{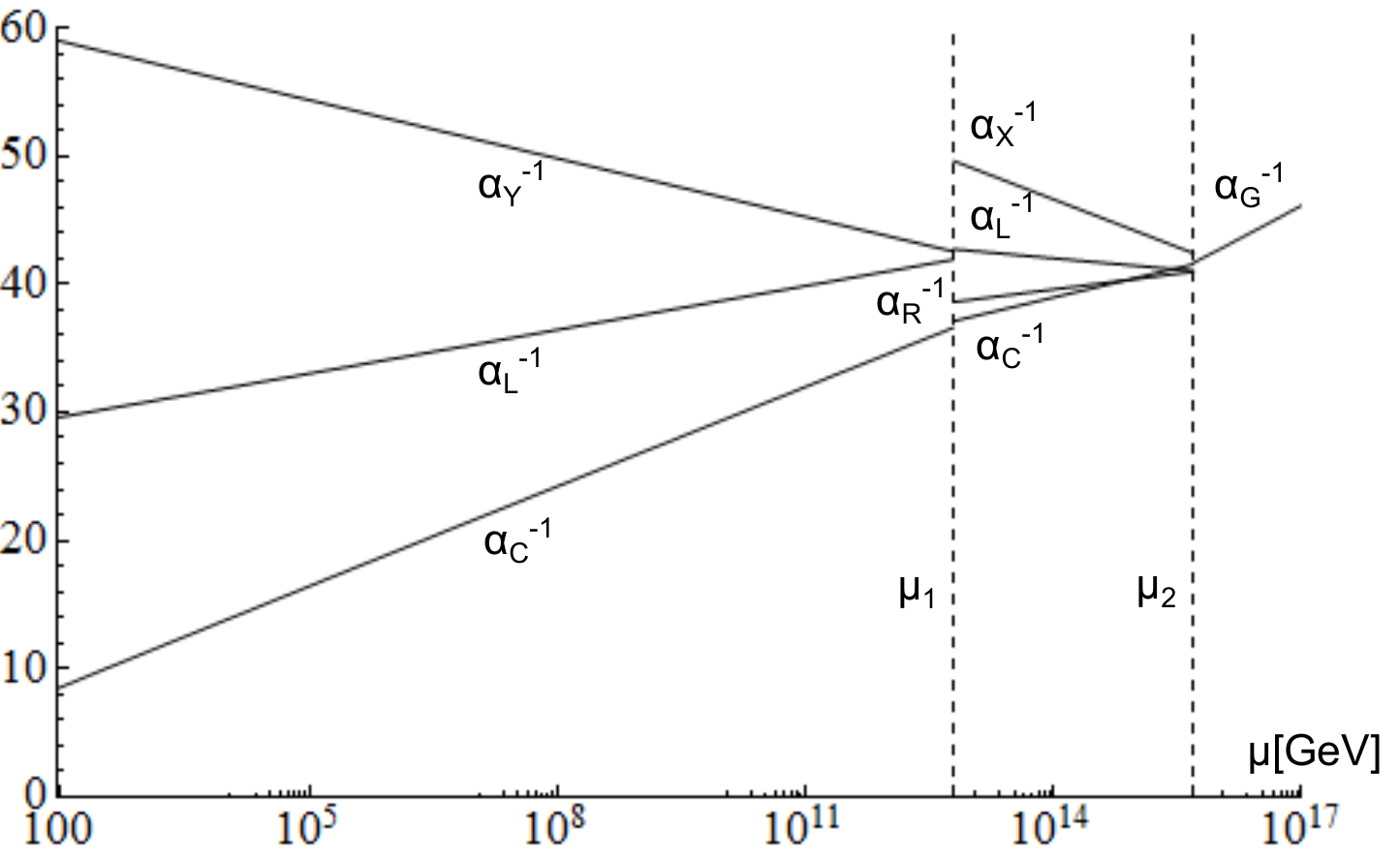}
\end{center}
\caption{The same as in FIG.~\ref{FigRun1} but for Point 2 in TABLE~\ref{TabPar}.}
\label{FigRun2}
\end{figure}

\subsubsection{Further remarks}
\paragraph{Two loop effects of $M(6,3,+\frac{1}{3})$.} \label{parM6} It is well known that at the NLO level the two-loop running effects are generally comparable to the one-loop threshold corrections if the fields that are integrated out cluster around the matching scale. Since, however, the fine-tuned mass of the $(6,3,+\frac{1}{3})$ scalar ``slides'' from the vicinity of $\mu_1$ where it is integrated out (see the dashed line at Figure \ref{FigM6Sigma}) to as high as  $\mu_2$ there is a danger that the threshold effects in the latter case can become larger than expected  (see Figure \ref{FigRun1}) and, hence, the hierarchy of the corrections may get out of control. To this end, we checked  the consistency of our calculation in the most extreme cases by introducing yet another  matching scale at the very sextet mass $M(6,3,+\frac{1}{3})$ and considering different effective theories above and below this threshold; in such setting, the size of the possible deviation from the simplified treatment should  mimic the possibly large two-loop threshold corrections. However, numerically, these effects turn out to be very small; the reason is that the changes of the inclinations of the parameter space boundaries in FIGs. \ref{FigM6BLShrink} and \ref{FigM6Sigma} play, to a large extent, against each other and, thus, the shape of the essential $\omega_{BL}$-$|\sigma|$ correlation remains intact.

\paragraph{Choice of the matching scales.\label{matchingindependence}} The consistency of the entire treatment of the threshold corrections can be checked rather easily by recalling that the $n$-th loop thresholds should make the effective SM gauge couplings independent of the choice of the matching scale up to the same level. Hence, the possible residual dependence of our results on the choice of the matching scales should correspond to the two-loop thresholds which, in size, are comparable to three-loop $\beta$-function effects. We checked this behavior for each of the consistent points; the typical change in the low-scale couplings inflicted by, e.g., increasing $\mu_1$ by a factor of 3 leads to shifts of the order of $10^{-2}$ in $\alpha_i^{-1}(M_Z)$ which, indeed, is in the right ballpark of a typical two-loop threshold/three-loop $\beta$-function effect. Besides that, these effects are comparable to the uncertainties in the input data~\eqref{AlphaErrors}.

\paragraph{Effects of the $10_H$ representation.}
\label{Par10H}
As we mentioned above, realistic fermion masses may be obtained only if the Higgs sector contains at least one more ``Yukawa-active'' representation besides $\Sigma$. The most simple and popular choice is then the ten-dimensional vector;  its $(1,2,2,0)$ part (in the $3_c2_L2_R1_{BL}$ notation) mixes with the same multiplet from $\Sigma$ below the Pati-Salam breaking scale. The mass matrix of the resulting four $SU(2)_{L}$ doublets must be fine-tuned so that one of them becomes the SM Higgs with an electroweak-scale effective mass parameter. Hence, an entire $(1,2,2,0)$ multiplet (i.e., an appropriate mixture of the two relevant fields) must survive down to the $SU(2)_R$ symmetry breaking scale. Since we work in a slightly simplified setting without the extra scalar $10$ at play we just mimic this situation by putting the next-to-lightest mass eigenvalue of the $SU(2)_{L}$-doublet mass matrix to the $SU(2)_R$-breaking scale, i.e., $M(1,2,+\frac{1}{2})=M(1,1,\pm1)_{VB}$ and assume that the remaining fields in the $(1,2,2,0)$ sector are integrated out at exactly the GUT scale and, as such, their possible threshold effects would be sub-dominant and, hence, leave our results intact.

\paragraph{BBN constraints.}
\label{ParBBN}
Unlike for the setting with the TeV-scale octet whose late decays may be, in principle, dangerous for the BBN (but only if its Yukawa couplings happen to be significantly suppressed) there is hardly any concern like this in the sextet case; indeed, such a ``light exotics'' here is so heavy that its natural decay width is parametrically different from that of the octet which, in turn, makes the sextet scenario very safe in this respect. 

%%%%%%%%%%%%%%%%%%%%%%%%%%%%%%%%%%%%%%
\section{Conclusions and outlook\label{conclusionsoutlook}}
%%%%%%%%%%%%%%%%%%%%%%%%%%%%%%%%%%%%%%
In this work we have recapitulated in detail the structure and the current status of the minimal potentially realistic renormalizable $SO(10)$ grand unified model with the high-scale gauge symmetry broken by the adjoint representation plus a single copy of the five-index fully antisymmetric self-dual tensor. Unlike for most of its alternatives (e.g., models with either 54 or 210 in the scalar sector responsible for the GUT symmetry breaking), the absence of the leading Planck-suppressed $d=5$ correction to the GUT-scale gauge kinetic form makes this setting  very robust with respect to the quantum gravity effects; this, in turn, makes it particularly suitable for the precision proton lifetime calculations. 
Indeed, in the current scenario, the 
scale of the perturbative baryon and lepton number violation (i.e., the GUT scale), may be, in principle, reliably calculated to the two-loop order in the perturbation expansion. Consequently, the corresponding theoretical uncertainties in the proton lifetime estimates are expected to be under much better control than in other models and even comparable to the size of the sensitivity improvement window of the upcoming megaton-scale experiments such as the Hyper-Kamiokande.       

In particular, we attempted to conclude the first step in this program, which is the detailed two-loop determination of the parameter space compatible with the basic phenomenological constraints (namely, those coming from the gauge unification and proton lifetime) paying particular attention to the overall dynamical consistency of the picture. This, in fact, can be attained only at the {\em quantum} level due to the severe tachyonic instabilities developing in the tree-level spectrum along the physically interesting symmetry breaking chains. Hence, the NLO approach to the minimal $SO(10)$ model under consideration is a must rather than an option.

At the vast majority of the parameter space there turn out to be just two classes of solutions conforming all the requirements specified in Sect.~\ref{sect:consistency}. The first of them, studied in great detail in the recent work~\cite{Bertolini:2013vta}, is a light color octet with hypercharge $\tfrac{1}{2}$ transforming as a weak isospin doublet with mass below about 20 TeV while the second option consists in having an intermediate-scale color sextet with hypercharge $\tfrac{1}{3}$ transforming like an $SU(2)_{L}$ triplet. Concerning the former, this class of solutions is very interesting due to a clear anti-correlation between the octet mass and the proton lifetime; remarkably enough, this relation turns out to be so tight that {\em it either implies the octet to be visible at the LHC or one of its near future successors or the proton decay to be observable at Hyper-K} (assuming it reaches its design sensitivity).  As for the sextet, the NLO analysis also reveals an interesting though slightly more complicated correlation among the proton longevity, the mass of the sextet and the absolute neutrino mass scale (barring possible multiply-finetuned settings with a strongly suppressed Dirac neutrino mass matrix) which leaves only a very little room for the sextet at around $10^{14}$~GeV if proton decay would not be seen at Hyper-K.

There is a further comment worth at this point: Although the existing $SO(10)$ renormalizable Yukawa sector fits (such as~\cite{Joshipura:2011nn,Dueck:2013gca}) do not admit $\sigma$ below about $10^{12}$ GeV, one should refrain from arguing that the sextet solution would be essentially ruled-out if there was no $p$-decay seen at the Hyper-K. This is namely due to the fact that these fits were done under the simplifying assumption that there are just two matrices governing the Yukawa sector of the model 
(as it is the case, for instance, in the minimal supersymmetric SO(10) GUT~\cite{Clark:1982ai,Aulakh:2003kg}). Let us remark that, in the non-SUSY case,  this is a strong extra assumption because there are two possible contractions of the $10_{S}$ with matter ($16_{M}16_{M}10_{S}$ and/or $16_{M}16_{M}10_{S}^{*}$) allowed due to the reality of the SO(10) vector representation; hence, the most general renormalizable Yukawa Lagrangian in the non-SUSY models with $10_{S}\oplus 126_{S}$ is governed by three rather than two independent complex symmetric matrices, cf.~\cite{Bajc:2005zf}. 
From this point of view, the simplified setting calls for further justification; this is often done by invoking an extra global symmetry of the Peccei-Quinn (PQ) type which forbids one of the two couplings and, at the same time, serves as a means to resolve the SM strong CP problem and provides an invisible axion as a dark matter candidate. However, this general scheme is not easily implemented in the model under consideration because, without extra structure, there is always a global remnant of the original PQ symmetry surviving down to the electroweak scale, in conflict with the current axion bounds.  

Hence, one should take the quoted lower bounds on $\sigma$ with a grain of salt as, in the most general case, the Yukawa sector may be capable of accommodating an arbitrarily small Dirac neutrino mass matrix without trouble with the charged sector fits and, hence, yield acceptable light neutrino masses even for $\sigma$ much below its ``natural'' domain at around $10^{12-14}$ GeV. Unfortunately, this issue can be settled only by a dedicated numerical analysis of the most general setting which, however, is out of the scope of the current study.

\section*{Acknowledgments}
The work of M.M. is supported by the Marie-Curie Career Integration Grant within the 7th European Community Framework Programme
FP7-PEOPLE-2011-CIG, contract number PCIG10-GA-2011-303565, by the Charles University grant PRVOUK P45, by the Foundation for support of science and research ``Neuron'' and by the Research proposal MSM0021620859 of the Ministry of Education, Youth and Sports of the Czech Republic. The work of H.K. is supported by the Grant Agency of the Czech Technical University in Prague, grant No. SGS13/217/OHK4/3T/14. 
We are indebted to Stefano Bertolini and Luca Di Luzio for initial discussions and invaluable comments.

\appendix

%%%%%%%%%%%%%%%%%%%%%%%%%%%
\section*{Appendix}
%%%%%%%%%%%%%%%%%%%%%%%%%%%
\section{One-loop matching\label{sect:matching}}
\label{ApMatching}
In this Appendix we list the explicit forms of all the matching functions defined in Section~\ref{sect:running}. In order to simplify the notation, those related to different stages of the RG evolution of the same gauge factor are denoted by the same functional symbol; their association to a specific matching is indicated by the corresponding matching-scale variables. 
\subsection{The $SO(10) \to 3_c 2_L 2_R 1_X$ matching at $\mu_2$\label{app:mu2}}
The threshold functions associated to the ``SM gauge factors'' $SU(3)_{c}\otimes SU(2)_{L}$ entering the formulae (\ref{SU3ccondatmu2})-(\ref{SU2Lcondatmu2}) read
\begin{widetext}
\begin{align*}
\lambda_c(\mu_2) &=\frac{5}{48\pi^2}+
\frac{1}{8\pi^2} \Biggl[ -\frac{22}{3}\left(\log\frac{M(3,2,+1/6)_{VB}}{\mu_2}+\log\frac{M(3,2,-5/6)_{VB}}{\mu_2}\right)\\
&-\frac{11}{3}\log\frac{M(3,1,+2/3)_{VB}}{\mu_2}
+ \frac{1}{3} \left(\log\frac{M(3,2,+1/6)_{GB}}{\mu_2}+\log\frac{M(3,2,-5/6)_{GB}}{\mu_2}\right)\\
&+ \frac{1}{6}\log\frac{M(3,1,+2/3)_{GB}}{\mu_2}\\
&+\frac{1}{3}\left(\log\frac{M(3,2,+1/6)_{CS}^{(2)}}{\mu_2} + \log\frac{M(3,2,+1/6)_{CS}^{(3)}}{\mu_2}
+ \log\frac{M(3,2,+7/6)_{CS}^{(1)}}{\mu_2} + \log\frac{M(3,2,+7/6)_{CS}^{(2)}}{\mu_2}\right)\\
&+ \frac{1}{6}\left(\log\frac{M(3,1,-4/3)_{CS}}{\mu_2} + \log\frac{M(3,1,-1/3)_{CS}^{(1)}}{\mu_2}+ \log\frac{M(3,1,+2/3)_{CS}^{(2)}}{\mu_2}\right)\\
&+\frac{1}{6}\left(\log\frac{M(3,1,-1/3)_{CS}^{(2)}}{\mu_2} + \log\frac{M(3,1,-1/3)_{CS}^{(3)}}{\mu_2}\right)
+\frac{1}{2} \log\frac{M(\bar{3},3,+1/3)_{CS}}{\mu_2}\\
&+\frac{5}{6}\left(\log\frac{M(6,1,+4/3)_{CS}}{\mu_2}+\log\frac{M(6,1,+1/3)_{CS}}{\mu_2} + \log\frac{M(6,1,-2/3)_{CS}}{\mu_2}\right)\\
&+ \frac{1}{2}\log\frac{M(8,1,0)_{RS}}{\mu_2}
+ 2\left(\log\frac{M(8,2,+1/2)_{CS}^{(1)}}{\mu_2}+\log\frac{M(8,2,+1/2)_{CS}^{(2)}}{\mu_2}\right)\Biggr]
\end{align*}
\begin{align*}
\lambda_L(\mu_2)&= \frac{1}{8\pi^2} +\frac{1}{8\pi^2}\Biggl[
-11\left(\log\frac{M(3,2,+1/6)_{VB}}{\mu_2}+\log\frac{M(3,2,-5/6)_{VB}}{\mu_2}\right) \\
&+ \frac{1}{2} \left(\log\frac{M(3,2,+1/6)_{GB}}{\mu_2}+\log\frac{M(3,2,-5/6)_{GB}}{\mu_2}\right)\\
&+\frac{1}{2}\left(\log\frac{M(3,2,+1/6)_{CS}^{(2)}}{\mu_2} + \log\frac{M(3,2,+1/6)_{CS}^{(3)}}{\mu_2}
+ \log\frac{M(3,2,+7/6)_{CS}^{(1)}}{\mu_2} + \log\frac{M(3,2,+7/6)_{CS}^{(2)}}{\mu_2}\right)\\
&+ 2 \log\frac{M(\bar{3},3,+1/3)_{CS}}{\mu_2}+ \frac{4}{3}\left(\log\frac{M(8,2,+1/2)_{CS}^{(1)}}{\mu_2}+\log\frac{M(8,2,+1/2)_{CS}^{(2)}}{\mu_2}\right)\\
&+\frac{1}{3}\log\frac{M(1,3,0)_{RS}}{\mu_2} + \frac{2}{3}\log\frac{M(1,3,+1)_{CS}}{\mu_2}
 \Biggr]
\end{align*}
\end{widetext}
Let us reiterate that due to the $3_c 2_L 1_R 1_X$ classification of the ``heavy'' spectrum it is quite convenient to introduce a matrix notation for the $SU(2)_{R}\otimes U(1)_{X}$ couplings already at this level, cf. Sect.~\ref{SecThreshold}. The elements of the relevant ``threshold matrix'' $\Lambda$ (\ref{matLambda}) then read 
\begin{widetext}
\begin{align*}
\lambda_{RR}(\mu_2)&=\frac{1}{8\pi^2} + \frac{1}{8\pi^2}\Biggl[  -11\left(\log\frac{M(3,2,+1/6)_{VB}}{\mu_2}+\log\frac{M(3,2,-5/6)_{VB}}{\mu_2}\right)\\
&+ \frac{1}{2} \left(\log\frac{M(3,2,+1/6)_{GB}}{\mu_2}+\log\frac{M(3,2,-5/6)_{GB}}{\mu_2}\right)\\
&+\frac{1}{2}\left(\log\frac{M(3,2,+1/6)_{CS}^{(2)}}{\mu_2} + \log\frac{M(3,2,+1/6)_{CS}^{(3)}}{\mu_2}
+ \log\frac{M(3,2,+7/6)_{CS}^{(1)}}{\mu_2} + \log\frac{M(3,2,+7/6)_{CS}^{(2)}}{\mu_2}\right)\\
&+ \left(\log\frac{M(3,1,-4/3)_{CS}}{\mu_2} + \log\frac{M(3,1,+2/3)_{CS}^{(2)}}{\mu_2}\right)+ 2 \left(\log\frac{M(6,1,+4/3)_{CS}}{\mu_2} + \log\frac{M(6,1,-2/3)_{CS}}{\mu_2}\right)\\
&+ \frac{4}{3}\left(\log\frac{M(8,2,+1/2)_{CS}^{(1)}}{\mu_2}+\log\frac{M(8,2,+1/2)_{CS}^{(2)}}{\mu_2}\right)+\frac{1}{3}\log\frac{M(1,1,+1)_{CS}^{(2)}}{\mu_2} \Biggr]
\end{align*}
\begin{align*}
\lambda_{XX}(\mu_2)&= \frac{1}{6\pi^2}+\frac{1}{8\pi^2} \Biggl[ -\frac{22}{3}\left(\log\frac{M(3,2,+1/6)_{VB}}{\mu_2}+\log\frac{M(3,2,-5/6)_{VB}}{\mu_2}\right) -\frac{44}{3}\log\frac{M(3,1,+2/3)_{VB}}{\mu_2}\\
&+ \frac{1}{3} \left(\log\frac{M(3,2,+1/6)_{GB}}{\mu_2}+\log\frac{M(3,2,-5/6)_{GB}}{\mu_2}\right)+ \frac{2}{3}\log\frac{M(3,1,+2/3)_{GB}}{\mu_2}\\
&+\frac{4}{3}\left(\log\frac{M(3,2,+1/6)_{CS}^{(2)}}{\mu_2} + \log\frac{M(3,2,+1/6)_{CS}^{(3)}}{\mu_2}
+ \log\frac{M(3,2,+7/6)_{CS}^{(1)}}{\mu_2} + \log\frac{M(3,2,+7/6)_{CS}^{(2)}}{\mu_2}\right)\\
&+ \frac{1}{6}\left(\log\frac{M(3,1,-4/3)_{CS}}{\mu_2} + \log\frac{M(3,1,-1/3)_{CS}^{(1)}}{\mu_2}+ \log\frac{M(3,1,+2/3)_{CS}^{(2)}}{\mu_2}\right)\\
&+\frac{1}{6}\left(\log\frac{M(3,1,-1/3)_{CS}^{(2)}}{\mu_2} + \log\frac{M(3,1,-1/3)_{CS}^{(3)}}{\mu_2}\right)
+\frac{1}{2} \log\frac{M(\bar{3},3,+1/3)_{CS}}{\mu_2}\\
&+\frac{1}{3}\left(\log\frac{M(6,1,+4/3)_{CS}}{\mu_2}+\log\frac{M(6,1,+1/3)_{CS}}{\mu_2} + \log\frac{M(6,1,-2/3)_{CS}}{\mu_2}\right)+\frac{3}{2}\log\frac{M(1,3,+1)_{CS}}{\mu_2}\Biggr]
\end{align*}
\begin{align*}
\lambda_{RX}(\mu_2)&=\lambda_{XR}(\mu_2)=\frac{1}{8\pi^2}\sqrt{\frac{3}{8}}\Biggl[\frac{44}{3} \log\frac{M(3,2,+1/6)_{VB}}{M(3,2,-5/6)_{VB}} - \frac{2}{3} \log\frac{M(3,2,+1/6)_{GB}}{M(3,2,-5/6)_{GB}}\\
&+\frac{4}{3}\left(\log\frac{M(3,2,+7/6)_{CS}^{(1)}}{M(3,2,+1/6)_{CS}^{(2)}} + \log\frac{M(3,2,+7/6)_{CS}^{(2)}}{M(3,2,+1/6)_{CS}^{(3)}}\right)- \frac{2}{3} \log\frac{M(3,1,+2/3)_{CS}^{(2)}}{M(3,1,-4/3)_{CS}} + \frac{4}{3} \log\frac{M(6,1,+4/3)_{CS}}{M(6,1,-2/3)_{CS}}\Biggr]
\end{align*}
\end{widetext}
Note that the off-diagonal $\lambda_{RX,XR}$ factors are, indeed, $\mu_{2}$-independent which reflects their auxiliary role; technically, this feature is implied by the zero trace of the $SU(2)_{R}$ generators.  
\subsection{\label{app:matchingmu1prime}The $3_c 2_L 2_R 1_X \to 3_c 2_L 1_R 1_X $ matching at $\mu_1'$ 
}
In the settings where there is a clearly identifiable $3_c 2_L 2_R 1_X$ stage, i.e., for those with $|\omega_{R}|\gg|\sigma|$ an extra matching scale $\mu_{1}'$ was conveniently introduced (cf. Sects.~\ref{efftheories} and~\ref{SecThreshold}).  The relevant threshold factors are then given by
\begin{widetext}
 \begin{align*}
    \lambda_c(\mu_1')&=0\\
    \lambda_L(\mu_1')&=\frac{1}{48\pi^2}\log\frac{M(1,2,+1/2)_{CS}}{\mu_1'}\\
    \lambda_{RR}(\mu_1')&=\frac{1}{24\pi^2} -\frac{22}{24\pi^2}\log\frac{M(1,1,+1)_{VB}}{\mu_1'} + \frac{1}{48\pi^2}\log\frac{M(1,2,+1/2)_{CS}}{\mu_1'}+ \frac{1}{24\pi^2}\log\frac{M(1,1,-2)_{CS}}{\mu_1'}\\
    \lambda_{RX}(\mu_1')&=\lambda_{XR}(\mu_1')=\frac{\sqrt{6}}{48\pi^2}\log\frac{M(1,1,-2)_{CS}}{\mu_1'}\\
    \lambda_{XX}(\mu_1')&=\frac{1}{16\pi^2}\log\frac{M(1,1,+1)_{GB}}{\mu_1'}+\frac{1}{16\pi^2}\log\frac{M(1,1,-2)_{CS}}{\mu_1'}
\end{align*}
\end{widetext}
\subsection{Matching to the SM at $\mu_1$}
The specific shape of the threshold factors relevant for the final matching of the relevant effective gauge theory to the SM  depends on the whether there is an intermediate $3_c 2_L 1_R 1_X$ symmetry encountered along the relevant breaking chain or not. 
\subsubsection{Matching $3_c 2_L 2_R 1_X$ to the SM at $\mu_1$ -  chain (\ref{BreakingSimple})\label{app:mu1matchingA}}
If it is not the case, i.e., if $|\omega_{R}|<|\sigma|$ and it is reasonable to reduce the intermediate $3_c 2_L 2_R 1_X$ symmetry right to the SM, the relevant threshold functions are given by
\begin{widetext}
\begin{align*}
    \lambda_c(\mu_1)&=\frac{5}{16\pi^2}\log\frac{M(6,3,+1/3)_{CS}}{\mu_1}\\
    \lambda_L(\mu_1)&=\frac{1}{48\pi^2}\log\frac{M(1,2,+1/2)_{CS}}{\mu_1}+\frac{1}{2\pi^2}\log\frac{M(6,3,+1/3)_{CS}}{\mu_1}\\
    \lambda_{Y}(\mu_1)&=\frac{1}{40\pi^2} -\frac{11}{20\pi^2}\log\frac{M(1,1,+1)_{VB}}{\mu_1} + \frac{1}{80\pi^2}\log\frac{M(1,2,+1/2)_{CS}}{\mu_1}+ \frac{1}{10\pi^2}\log\frac{M(1,1,-2)_{CS}}{\mu_1} \\ &+\frac{1}{20\pi^2}\log\frac{M(6,3,+1/3)_{CS}}{\mu_1}
\end{align*}

\end{widetext}
\subsubsection{Matching $3_c 2_L 1_R 1_X$ to the SM at $\mu_1$ -  chain (\ref{BreakingComplicated})}
\label{app:mu1matchingB}
In this mode the situation is even simpler that in the previous case because, besides the full SM singlets (that do not contribute to the matching factors at all)  there is  only the sextet to be integrated out at $\mu_1$; the relevant formulae read  
\begin{align*}
    \lambda_{c}(\mu_1)&=\frac{5}{16\pi^2}\log\frac{M(6,3,+1/3)_{CS}}{\mu_1}\\
    \lambda_{L}(\mu_1)&=\frac{1}{2\pi^2}\log\frac{M(6,3,+1/3)_{CS}}{\mu_1}\\
    \lambda_{Y}(\mu_1)&=\frac{1}{20\pi^2}\log\frac{M(6,3,+1/3)_{CS}}{\mu_1}.
\end{align*}
\section{Sample scalar spectrum}
\label{ApSpectrum}
In TABLE~\ref{TabSpec1} we present the spectrum of the heavy vectors and scalars for the two sample points specified in TABLE~\ref{TabPar}. The relevant fields are classified with respect to the SM gauge group. Notice the difference in the position of the $\left(6,3,+\frac{1}{3}\right)$ scalar field and in the masses of the vector bosons associated to different symmetry breaking scales (in particular, the $\left(3,2,-\frac{5}{6}\right)$ and $\left(3,2,+\frac{1}{6}\right)$ vectors responsible for the $d=6$ proton decay) between the left and right panels. Notice also that for both points one has $|\omega_R|<|\sigma|$ and, thus, the mass of the $(1,1,\pm1)_{VB}$ vector boson is lower than that of $(1,1,0)_{VB}$.

\begin{widetext}
\begin{center}
\setlength{\tabcolsep}{10pt}
\begin{table}%[h]
\begin{tabular}{c c}
\def\arraystretch{1.4}
\setlength{\tabcolsep}{2pt}
\begin{tabular}{| >{$}c<{$} | >{$}c<{$} | >{$}c<{$} | >{$}c<{$} | >{$}c<{$} |}
\hline
\text{Multiplet X}&\text{Type}&\#&\Delta a_{X} & \text{Mass} \left[\mathrm{GeV}\right]\\
\hline
 \left(1,1,+1\right) & \text{VB} & 1 & \left(0,0,-\frac{11}{5}\right) & 9.2\times 10^{11} \\
 \left(1,1,-1\right) & \text{VB} & 1 & \left(0,0,-\frac{11}{5}\right) & 9.2\times 10^{11} \\
 \left(1,1,+1\right) & \text{GB} & 1 & \left(0,0,\frac{1}{5}\right) & 9.2\times 10^{11} \\
 \left(1,2,+\frac{1}{2}\right) & \text{CS} & 1 & \left(0,\frac{1}{12},\frac{1}{20}\right) & 9.2\times 10^{11} \\
  \left(1,1,0\right) & \text{VB} & 1 & \left(0,0,0\right) & 2.0\times 10^{12} \\
 \left(1,1,0\right) & \text{GB} & 1 & \left(0,0,0\right) & 2.0\times 10^{12} \\
 \left(6,3,+\frac{1}{3}\right) & \text{CS} & 1 & \left(\frac{5}{2},4,\frac{2}{5}\right) & 1.0\times 10^{15} \\
 \left(3,2,+\frac{1}{6}\right) & \text{CS} & 3 & \left(\frac{1}{3},\frac{1}{2},\frac{1}{30}\right) & 6.6\times 10^{15} \\
 \left(3,2,+\frac{7}{6}\right) & \text{CS} & 1 & \left(\frac{1}{3},\frac{1}{2},\frac{49}{30}\right) & 6.6\times 10^{15} \\
 \left(1,1,0\right) & \text{RS} & 2 & \left(0,0,0\right) & 6.8\times 10^{15} \\
 \left(1,1,-2\right) & \text{CS} & 1 & \left(0,0,\frac{4}{5}\right) & 6.8\times 10^{15} \\
 \left(1,1,+1\right) & \text{CS} & 2 & \left(0,0,\frac{1}{5}\right) & 6.8\times 10^{15} \\
 \left(1,1,0\right) & \text{RS} & 3 & \left(0,0,0\right) & 6.8\times 10^{15} \\
 \left(8,1,0\right) & \text{RS} & 1 & \left(\frac{1}{2},0,0\right) & 8.0\times 10^{15} \\
 \left(1,3,0\right) & \text{RS} & 1 & \left(0,\frac{1}{3},0\right) & 8.1\times 10^{15} \\
 \left(3,2,+\frac{1}{6}\right) & \text{VB} & 1 & \left(-\frac{11}{3},-\frac{11}{2},-\frac{11}{30}\right) & 9.0\times 10^{15} \\
 \left(\bar{3},2,-\frac{1}{6}\right) & \text{VB} & 1 & \left(-\frac{11}{3},-\frac{11}{2},-\frac{11}{30}\right) & 9.0\times 10^{15} \\ \left(3,2,+\frac{1}{6}\right) & \text{GB} & 1 & \left(\frac{1}{3},\frac{1}{2},\frac{1}{30}\right) & 9.0\times 10^{15} \\
 \left(3,2,-\frac{5}{6}\right) & \text{VB} & 1 & \left(-\frac{11}{3},-\frac{11}{2},-\frac{55}{6}\right) & 9.0\times 10^{15} \\
 \left(\bar{3},2,+\frac{5}{6}\right) & \text{VB} & 1 & \left(-\frac{11}{3},-\frac{11}{2},-\frac{55}{6}\right) & 9.0\times 10^{15} \\
 \left(3,2,-\frac{5}{6}\right) & \text{GB} & 1 & \left(\frac{1}{3},\frac{1}{2},\frac{5}{6}\right) &9.0\times 10^{15} \\
 \left(\bar{3},1,+\frac{1}{3}\right) & \text{CS} & 1 & \left(\frac{1}{6},0,\frac{1}{15}\right) & 1.1\times 10^{16} \\
 \left(8,2,+\frac{1}{2}\right) & \text{CS} & 1 & \left(2,\frac{4}{3},\frac{4}{5}\right) & 1.4\times 10^{16} \\
 \left(3,1,+\frac{2}{3}\right) & \text{VB} & 1 & \left(-\frac{11}{6},0,-\frac{44}{15}\right) & 1.8\times 10^{16} \\
 \left(\bar{3},1,-\frac{2}{3}\right) & \text{VB} & 1 & \left(-\frac{11}{6},0,-\frac{44}{15}\right) & 1.8\times 10^{16} \\
  \left(3,1,+\frac{2}{3}\right) & \text{GB} & 1 & \left(\frac{1}{6},0,\frac{4}{15}\right) & 1.8\times 10^{16} \\
 \left(1,1,0\right) & \text{RS} & 4 & \left(0,0,0\right) & 2.2\times 10^{16} \\
 \left(3,1,-\frac{4}{3}\right) & \text{CS} & 1 & \left(\frac{1}{6},0,\frac{16}{15}\right) & 2.2\times 10^{16} \\
 \left(\bar{3},1,+\frac{1}{3}\right) & \text{CS} & 2 & \left(\frac{1}{6},0,\frac{1}{15}\right) & 2.2\times 10^{16} \\
 \left(3,1,+\frac{2}{3}\right) & \text{CS} & 2 & \left(\frac{1}{6},0,\frac{4}{15}\right) & 2.2\times 10^{16} \\
 \left(8,2,+\frac{1}{2}\right) & \text{CS} & 2 & \left(2,\frac{4}{3},\frac{4}{5}\right) & 2.3\times 10^{16} \\
 \left(6,1,-\frac{2}{3}\right) & \text{CS} & 1 & \left(\frac{5}{6},0,\frac{8}{15}\right) & 2.6\times 10^{16} \\
 \left(6,1,+\frac{1}{3}\right) & \text{CS} & 1 & \left(\frac{5}{6},0,\frac{2}{15}\right) & 2.6\times 10^{16} \\
 \left(6,1,+\frac{4}{3}\right) & \text{CS} & 1 & \left(\frac{5}{6},0,\frac{32}{15}\right) & 2.6\times 10^{16} \\
 \left(\bar{3},1,+\frac{1}{3}\right) & \text{CS} & 3 & \left(\frac{1}{6},0,\frac{1}{15}\right) & 3.2\times 10^{16} \\
 \left(\bar{3},3,+\frac{1}{3}\right) & \text{CS} & 1 & \left(\frac{1}{2},2,\frac{1}{5}\right) & 3.4\times 10^{16} \\
 \left(3,2,+\frac{1}{6}\right) & \text{CS} & 2 & \left(\frac{1}{3},\frac{1}{2},\frac{1}{30}\right) & 3.7\times 10^{16} \\
 \left(3,2,+\frac{7}{6}\right) & \text{CS} & 2 & \left(\frac{1}{3},\frac{1}{2},\frac{49}{30}\right) & 3.7\times 10^{16} \\
 \left(1,3,+1\right) & \text{CS} & 1 & \left(0,\frac{2}{3},\frac{3}{5}\right) & 4.5\times 10^{16}\\
 \hline
\end{tabular}
&
\def\arraystretch{1.4}
\setlength{\tabcolsep}{2pt}
\begin{tabular}{| >{$}c<{$} | >{$}c<{$} | >{$}c<{$} | >{$}c<{$} | >{$}c<{$} |}
\hline
\text{Multiplet X}&\text{Type}& \# &\Delta a_{X} & \text{Mass} \left[\mathrm{GeV}\right]\\
\hline
 \left(1,1,+1\right) & \text{VB} & 1 & \left(0,0,-\frac{11}{5}\right) & 9.8\times 10^{12} \\
 \left(1,1,-1\right) & \text{VB} & 1 & \left(0,0,-\frac{11}{5}\right) & 9.8\times 10^{12} \\
 \left(1,1,+1\right) & \text{GB} & 1 & \left(0,0,\frac{1}{5}\right) & 9.8\times 10^{12} \\
 \left(1,2,+\frac{1}{2}\right) & \text{CS} & 1 & \left(0,\frac{1}{12},\frac{1}{20}\right) & 9.8\times 10^{12} \\
 \left(1,1,0\right) & \text{VB} & 1 & \left(0,0,0\right) & 1.9\times 10^{13} \\
  \left(1,1,0\right) & \text{GB} & 1 & \left(0,0,0\right) & 1.9\times 10^{13} \\
 \left(6,3,+\frac{1}{3}\right) & \text{CS} & 1 & \left(\frac{5}{2},4,\frac{2}{5}\right) & 2.2\times 10^{13} \\
 \left(1,1,0\right) & \text{RS} & 2 & \left(0,0,0\right) & 1.1\times 10^{15} \\
 \left(1,1,-2\right) & \text{CS} & 1 & \left(0,0,\frac{4}{5}\right) & 1.1\times 10^{15} \\
 \left(1,1,+1\right) & \text{CS} & 2 & \left(0,0,\frac{1}{5}\right) & 1.1\times 10^{15} \\
 \left(1,1,0\right) & \text{RS} & 3 & \left(0,0,0\right) & 1.1\times 10^{15} \\
 \left(\bar{3},1,+\frac{1}{3}\right) & \text{CS} & 1 & \left(\frac{1}{6},0,\frac{1}{15}\right) & 1.9\times 10^{15} \\
 \left(8,1,0\right) & \text{RS} & 1 & \left(\frac{1}{2},0,0\right) & 2.6\times 10^{15} \\
 \left(1,3,0\right) & \text{RS} & 1 & \left(0,\frac{1}{3},0\right) & 2.7\times 10^{15} \\
 \left(3,2,+\frac{1}{6}\right) & \text{CS} & 3 & \left(\frac{1}{3},\frac{1}{2},\frac{1}{30}\right) & 3.0\times 10^{15} \\
 \left(3,2,+\frac{7}{6}\right) & \text{CS} & 1 & \left(\frac{1}{3},\frac{1}{2},\frac{49}{30}\right) & 3.0\times 10^{15} \\
 \left(8,2,+\frac{1}{2}\right) & \text{CS} & 1 & \left(2,\frac{4}{3},\frac{4}{5}\right) & 3.4\times 10^{15} \\
 \left(3,2,+\frac{1}{6}\right) & \text{VB} & 1 & \left(-\frac{11}{3},-\frac{11}{2},-\frac{11}{30}\right) & 4.9\times 10^{15} \\
 \left(\bar{3},2,-\frac{1}{6}\right) & \text{VB} & 1 & \left(-\frac{11}{3},-\frac{11}{2},-\frac{11}{30}\right) & 4.9\times 10^{15} \\ \left(3,2,+\frac{1}{6}\right) & \text{GB} & 1 & \left(\frac{1}{3},\frac{1}{2},\frac{1}{30}\right) & 4.9\times 10^{15} \\
 \left(3,2,-\frac{5}{6}\right) & \text{VB} & 1 & \left(-\frac{11}{3},-\frac{11}{2},-\frac{55}{6}\right) & 4.9\times 10^{15} \\
 \left(\bar{3},2,+\frac{5}{6}\right) & \text{VB} & 1 & \left(-\frac{11}{3},-\frac{11}{2},-\frac{55}{6}\right) & 4.9\times 10^{15} \\
  \left(3,2,-\frac{5}{6}\right) & \text{GB} & 1 & \left(\frac{1}{3},\frac{1}{2},\frac{5}{6}\right) & 4.9\times 10^{15} \\
 \left(6,1,-\frac{2}{3}\right) & \text{CS} & 1 & \left(\frac{5}{6},0,\frac{8}{15}\right) & 7.5\times 10^{15} \\
 \left(6,1,+\frac{1}{3}\right) & \text{CS} & 1 & \left(\frac{5}{6},0,\frac{2}{15}\right) & 7.5\times 10^{15} \\
 \left(6,1,+\frac{4}{3}\right) & \text{CS} & 1 & \left(\frac{5}{6},0,\frac{32}{15}\right) & 7.5\times 10^{15} \\
 \left(8,2,+\frac{1}{2}\right) & \text{CS} & 2 & \left(2,\frac{4}{3},\frac{4}{5}\right) & 9.1\times 10^{15} \\
 \left(3,1,+\frac{2}{3}\right) & \text{VB} & 1 & \left(-\frac{11}{6},0,-\frac{44}{15}\right) & 9.8\times 10^{15} \\
 \left(\bar{3},1,-\frac{2}{3}\right) & \text{VB} & 1 & \left(-\frac{11}{6},0,-\frac{44}{15}\right) & 9.8\times 10^{15} \\
 \left(3,1,+\frac{2}{3}\right) & \text{GB} & 1 & \left(\frac{1}{6},0,\frac{4}{15}\right) & 9.8\times 10^{15} \\
 \left(3,1,-\frac{4}{3}\right) & \text{CS} & 1 & \left(\frac{1}{6},0,\frac{16}{15}\right) & 1.0\times 10^{16} \\
 \left(\bar{3},1,+\frac{1}{3}\right) & \text{CS} & 2 & \left(\frac{1}{6},0,\frac{1}{15}\right) & 1.0\times 10^{16} \\
 \left(3,1,+\frac{2}{3}\right) & \text{CS} & 2 & \left(\frac{1}{6},0,\frac{4}{15}\right) & 1.0\times 10^{16} \\
 \left(3,2,+\frac{1}{6}\right) & \text{CS} & 2 & \left(\frac{1}{3},\frac{1}{2},\frac{1}{30}\right) & 1.2\times 10^{16} \\
 \left(3,2,+\frac{7}{6}\right) & \text{CS} & 2 & \left(\frac{1}{3},\frac{1}{2},\frac{49}{30}\right) & 1.2\times 10^{16} \\
 \left(\bar{3},1,+\frac{1}{3}\right) & \text{CS} & 3 & \left(\frac{1}{6},0,\frac{1}{15}\right) & 1.3\times 10^{16} \\
 \left(\bar{3},3,+\frac{1}{3}\right) & \text{CS} & 1 & \left(\frac{1}{2},2,\frac{1}{5}\right) & 1.3\times 10^{16} \\
 \left(1,3,+1\right) & \text{CS} & 1 & \left(0,\frac{2}{3},\frac{3}{5}\right) & 1.3\times 10^{16} \\
 \left(1,1,0\right) & \text{RS} & 4 & \left(0,0,0\right) & 4.1\times 10^{16}\\
 \hline
\end{tabular}
\end{tabular}
\caption{\label{TabSpec1} The shape of the ``heavy spectrum'' of the model under consideration for the two sample points in the parameter space identified in TABLE~\ref{TabPar} (Point 1 on the left, Point 2 on the right hand side). For each field $X$ we quote its contribution to the one-loop $\beta$-function that, besides formula~(\ref{acoef}), enters also the evolution of the baryon-number violating $d=6$ operators of our interest, cf. Sect.~\ref{d6protondecay}.}
\end{table}
\end{center}
\end{widetext}

%\bibliographystyle{h-physrev5}
%\bibliography{bibliography}

\end{document}